\documentclass{JHEP3}
\usepackage[normalem]{ulem}
\usepackage{cite}
\usepackage{epsfig}
\usepackage{amsmath}
\usepackage{bm,longtable,enumerate}
\usepackage{epsfig}
\usepackage{amsmath}
\usepackage{mathrsfs}
\usepackage{amsthm}
\usepackage{amssymb}
\usepackage{oldgerm}
\usepackage{indentfirst}
\usepackage{setspace}
\usepackage{multicol}
\usepackage{multirow}
\def\beq{\begin{equation}}
\def\eeq{\end{equation}}
\def\beqa{\begin{eqnarray}}
\def\eeqa{\end{eqnarray}}

\title{Vector meson quasinormal modes in a finite-temperature AdS/QCD model}

\author{Luis A. H. Mamani$^{\ast,\,\dag,\,a}$, Alex S. Miranda$^{\ast,\,b}$,
Henrique Boschi-Filho$^{\ddag,\,c}$, and Nelson R. F. Braga$^{\ddag,\,d}$.
\\
$^{\ast}$Departamento de Ci\^encias Exatas e Tecnol\'ogicas,\\
Universidade Estadual de Santa Cruz,\\
Rodovia Jorge Amado, km 16, 45662-900, Ilh\'eus, BA, Brazil\\
$^{\dag}$Centro de Ci\^encias Naturais e Humanas,
Universidade Federal do ABC,\\
Rua Santa Ad\'elia 166, 09210-170, Santo Andr\'e, SP, Brazil\\
$^{\ddag}$Instituto de F\'{i}sica, Universidade
Federal do Rio de Janeiro,\\
Caixa Postal 68528, RJ 21941-972, Brazil\\
$^{a}$email: luis.mamani@ufabc.edu.br;
$^{b}$email: asmiranda@uesc.br;
$^{c}$email: boschi@if.ufrj.br;
$^{d}$email: braga@if.ufrj.br.}

\abstract{
We study the spectrum of vector mesons in a finite temperature plasma.
The plasma is holographically described by a black hole AdS/QCD model.
We compute the boundary retarded Green's function using AdS/CFT prescriptions.
The corresponding thermal spectral functions show quasiparticle peaks at low temperatures. 
Then we calculate the quasinormal modes of vector mesons in the soft-wall 
black hole geometry and analyse their temperature and momentum dependences.   }


\preprint{}

\begin{document}


\section{Introduction}
\label{introd}

Strong interactions are described by QCD. As it is well known, the coupling varies with the energy.
At high energies the coupling is small and the theory can be treated perturbatively 
giving rise to the asymptotic freedom. 
At low energies perturbation theory does not work and one needs alternative approaches. 
Gauge/string dualities provide an important tool to study non perturbative aspects of strong interactions.  

The connection between string and gauge theories started with the seminal work \cite{'tHooft:1973jz}
relating planar diagrams of non abelian gauge theories to string theory. 
More recently, a remarkable duality between ten-dimensional string theory or eleven-dimensional M-theory
and a conformal gauge theory on the corresponding spacetime boundary was found in \cite{Maldacena:1997re}.
This so called AdS/CFT correspondence relates, in particular, string theory in $\mbox{AdS}_{5}\times S^{5}$ 
space to four-dimensional $\mbox{SU}(N_{c})$ Yang-Mills gauge theory with large $N_{c}$
and extended ${\cal{N}}=4$ supersymmetry.
String theory at low energies is described by a supergravity theory. 
In this case, the AdS/CFT correspondence implies a gauge/gravity duality from which one can calculate 
correlation functions for gauge-field operators \cite{Witten:1998qj,Gubser:1998bc}. 

In this work we are interested in finite temperature properties of vector mesons. So, we will consider 
the finite temperature version of the AdS/CFT correspondence in the supergravity regime.
This is obtained by considering a black hole embedded in an AdS spacetime \cite{Witten:1998zw}.
A prescription to calculate retarded propagators at finite temperature in the gauge theory in
Minkowski space was found in \cite{Son:2002sd} (see also Refs. 
\cite{Herzog:2002pc,Skenderis:2008dh,Skenderis:2008dg}). This formulation involves purely
incoming-wave condition for the fields 
at the horizon, and such a condition represents the total absorption by the black hole
without any emission. 

In the AdS/CFT correspondence the gauge theory is conformal. So, to describe strong interactions 
one needs to break this symmetry. This is done in various phenomenological models known as AdS/QCD
where an infrared cut-off is introduced. An example is the hard-wall model which introduces
a hard cut-off on the bulk geometry \cite{BoschiFilho:2002ta,BoschiFilho:2002vd,deTeramond:2005su}. 
This model was also studied at finite temperature for example in \cite{BoschiFilho:2006pe}.

An alternative AdS/QCD model, that leads to linear Regge trajectories for vector 
mesons and glueballs \cite{Karch:2006pv,Colangelo:2007pt,Colangelo:2008us}, is the soft-wall
model. In this case one introduces a scalar field in the AdS geometry. This non uniform field
works as a smooth infrared cut-off for the dual gauge theory. The soft-wall model can also
be considered at finite temperature. In this case, there are two coexisting geometries,
with and without a black hole. At high temperatures the black hole geometry is (globally)
stable, while  the geometry without the black hole is stable for low temperatures.
The transition between these two regimes is a Hawking-Page phase transition\cite{Hawking:1982dh}
and was studied for the soft-wall model in \cite{Herzog:2006ra,BallonBayona:2007vp}.

We will study here the spectrum of vector mesons at finite temperature in the soft-wall model
considering, for all temperatures, a black hole embedded in the soft-wall background. As discussed
in \cite{Miranda:2009uw}, at intermediate and low temperatures this corresponds respectively to
the (supercooled) metastable and unstable phases of the plasma where the black hole is still present. 
Studying the supercooled phase of the plasma we will mimic the formation of vector mesons 
when the strong interacting plasma formed in heavy ion collisions cools down. 
At the high temperatures produced when the plasma is formed, there is initially only a deconfined phase.
During the cooling process the hadrons are formed.  
These two separate confined and deconfined phases have been studied using gauge/gravity duality. 
Here we study the transition between these two phases for the case of vector mesons.

The gravitational dual of the vector mesons is a gauge field in the soft-wall background. 
From the action of this gauge field we obtain the finite temperature retarded Green's
functions of vector meson operators using numerical and analytical techniques. 
The imaginary part of the retarded Green's function gives the spectral function. 
The spectral function in the soft-wall model was calculated for vector mesons in
\cite{Fujita:2009wc} and for scalar glueballs and scalar mesons in \cite{Colangelo:2009ra}. 
For spectral functions of other holographic models, see for example, 
Refs. \cite{Kovtun:2006pf,Teaney:2006nc, Peeters:2006iu, Myers:2008cj,Myers:2008me,Colangelo:2012jy}.

Then we study the quasinormal modes (QNM) of the bulk gauge field. These modes are identified with
the spectrum of vector mesons. We perform numerical analysis to obtain the real and imaginary parts of the
quasinormal frequencies. At high temperatures a good convergence is achieved using the power series 
method \cite{Horowitz:1999jd} while at low temperatures the Breit-Wigner resonance method
\cite{Berti:2009wx} works better. As expected, we find that the poles of the
Fourier-transformed retarded correlation functions correspond
to the black hole quasinormal (QN) frequencies. Quasinormal modes in the AdS/CFT correspondence
have been reviewed recently in Refs. \cite{Berti:2009kk, Konoplya:2011qq}.


\section{Vector mesons in the soft-wall model at finite temperature}
\label{model}

\subsection{The soft-wall model}

According to the AdS/CFT dictionary, normalizable solutions of the bulk fields are dual to
states of the boundary gauge theory. 
One assumes that this type of duality also holds for the AdS/QCD models.
In the soft-wall model the conformal invariance is broken by the introduction of an
infrared cut-off consisting of a background scalar field. 
This model was proposed in Ref. \cite{Karch:2006pv} to reproduce the approximate linear
Regge trajectories for vector mesons. It is defined at zero temperature in a five-dimensional
AdS spacetime, whose metric is given by 
\begin{equation}
ds^{2}\,=\,  g_{MN} dx^M dx^N \,=\,\frac{R^2}{\zeta^2} \left( \eta_{\mu\nu} dx^{\mu} dx^\nu  + d\zeta^2\right),
\label{AdS}
\end{equation}
\noindent where $\eta_{\mu\nu} = (-1,1,1,1) $ and $R$ is the AdS space radius. 
The action for a five dimensional gauge field in this model is
\begin{equation}
\label{action}
S=\frac{1}{4g_{5}^2}\int d^5 x \,\sqrt{-g} \, e^{-\Phi(\zeta)} \,F_{MN}F^{MN},
\end{equation}
\noindent where $g_{5}^2=16\pi^2 R/N_{c}^{2}$ and $\Phi(\zeta)$ is the background scalar field
with the form $\Phi(\zeta)=c \zeta^2$. This field plays the role of an infrared cut-off
where $\sqrt{c}\,$ represents a mass scale. This implies a discrete mass spectrum given by  
\begin{equation}
\label{Espectrum}
m_n^2 \,= \, 4c (n + 1) \, ,
\end{equation}
where $n$ is called radial quantum number \cite{Karch:2006pv}.
 
It is important to remark that the five dimensional background of the soft-wall 
model does not arise from a solution of the 5D Einstein equations. 
However it reproduces quite well the hadronic phenomenology, in particular the hadronic spectra.  
 
At finite temperature one replaces in the action the metric (\ref{AdS}) by the metric of
an asymptotically AdS black hole
\begin{equation}
\label{equation1}
ds^2 \, = \, \frac{R^2}{\zeta^2}\bigg[-f(\zeta )dt^2+dx^2+dy^2+dz^2\bigg]+\frac{R^2}{\zeta^2 f(\zeta)}
d\zeta^2,
\end{equation}
\noindent where $f(\zeta )=1-(\zeta/\zeta_{h})^4$. 
The coordinate $\zeta$ is defined in the interval $0\le \zeta \le \zeta_h$, 
$\zeta=0$ corresponds to the boundary of the space and $\zeta_{h}$ is the 
position of the horizon. The black hole temperature $T$ is given by
$T =1/ \pi \zeta_{h} $ which is also the temperature of the boundary field theory.
 
\subsection{Equations of motion}
    
We now study the equations of motion that come from the action (\ref{action})
with the metric (\ref{equation1}). We choose the radial gauge $A_\zeta = 0$ and,
without any loss of generality, look for plane wave solutions
of the form: $ A_\mu (\zeta , x, y, z, t ) = e^{-i\omega t + iq z } \,A_\mu ( \zeta , \omega , q) $
propagating in the $z $ direction with wave vector $k_\mu = (-\omega, 0, 0 , q)$.
The equations take the form
%
\begin{eqnarray}
\label{equation15}
&& \partial^2_{\zeta}A_{t}
-\left(\frac{1}{\zeta}+2c\,\zeta\right)
\partial_{\zeta}A_{t}
-\frac{q}{f}\left(q A_{t}+\omega A_{z}\right)=0,\\ \cr
      \label{equation16}
&&      \partial^2_{\zeta}A_{\alpha}+
      \left(\partial_{\zeta}\mbox{ln}f
      -\frac{1}{\zeta}-2c\,\zeta\right)
      \partial_{\zeta}A_{\alpha}+\frac{1}{f^2}\left(\omega^2-
      q^2f\right)A_{\alpha}=0, \;\;\;\; \quad (\alpha=x,y)
      \\ \cr
 \label{equation18}
 &&     \partial^2_{\zeta}A_{z}+
      \left(\partial_{\zeta}\mbox{ln}f
      -\frac{1}{\zeta}-2c\,\zeta\right)
       \partial_{\zeta}A_{z}+\frac{\omega}{f^2}
       \left(q A_{t}+\omega A_{z}\right)=0,
   \\ \cr
\label{equation19}
&&      \omega\partial_{\zeta}A_{t}+
      q f\partial_{\zeta}A_{z}=0.
      \end{eqnarray}
\noindent The corresponding equations for the electric field components  
$  E_{x} = \omega A_{x}\,,\,\, E_{y} = \omega A_{y}\,,$ \break
$ \,\, E_{z} = \omega A_{z}+q A_{t}\,,$
are
      \begin{equation}\label{equation25}
      \partial_{\zeta}^2E_{z}+
      \left(\frac{\omega^2\partial_{\zeta}\mbox{ln}f(\zeta)}{\omega^2
      -q^2f}-\frac{1}{\zeta}-2c\,\zeta\right)
      \partial_{\zeta}E_{z}+\frac{\omega^2-q^2f}{f^2}E_{z}=0,\hspace{1.9cm}
      \end{equation}
 
      \begin{equation}\label{equation26}
      \partial_{\zeta}^2 E_{\alpha}+\left(\partial_{\zeta}\mbox{ln}f
      -\frac{1}{\zeta}-2c\,\zeta\right)
       \partial_{\zeta}E_{\alpha}+\frac{\omega^2-q^2 f}{f^2}E_{\alpha}=0,
      \quad (\alpha=x,y).
      \end{equation}

\noindent Note that equation (\ref{equation25}) for the longitudinal component is singular for
$\zeta=0\,,\,\pm \zeta_h, $\break $\pm\left(\sqrt[4]{1-{\omega}^{2}/{q}^{2}}\right) \zeta_h$,
while equation (\ref{equation26}) for the transverse components is singular at
\break $\zeta = 0,\,\pm\zeta_h,\,\pm\infty$.
 
Now we perform a Bogoliubov transformation in the longitudinal component of the electric field: 
$E_z=e^{\mathscr{B}/2}\psi_{z}$, where  $\mathscr{B} = c \zeta^2 +\ln\left[\zeta(\omega^2-q^2f\,)/R\right]$.
We also introduce the tortoise coordinate  \cite{Miranda:2009uw,Horowitz:1999jd} defined as
    
      \begin{equation}\label{equation206}
       r_{*}=\frac{\zeta_h}{2}\bigg[-\arctan\left(\frac{\zeta}{\zeta_h}\right)
      +\frac{1}{2}\ln\bigg(\frac{\zeta_h-\zeta}{\zeta_h+\zeta}\bigg) \bigg],
      \end{equation}
such that $\partial_{r_{*}}=-f(\zeta)\partial_{\zeta}$. Then we obtain
a Schr\"odinger like equation for the transformed longitudinal component
      \begin{equation}\label{equation207}
       \partial_{r_{*}}^{2}\psi_{z}+\omega^{2}\psi_{z}=V_{_L}\, \psi_{z},
      \end{equation}
where the effective potential $V_{_L}$ is given by
      \begin{equation}\label{equation208}
       V_{_L}=q^{2}f+e^{\mathscr{B}/2}\,\partial_{r_{\ast}}^{2}e^{-\mathscr{B}/2}.
      \end{equation}

Following the same procedure for the transverse electric field components,
with the Bogoliubov transformation $E_\alpha = e^{\mathscr{A}/2}\psi_{\alpha}$
where $\mathscr{A}= c \zeta^2 + \ln(\zeta/R)$, we find the Schr\"odinger like equation 
      \begin{equation}\label{equation209}
       \partial_{r_{*}}^{2}\psi_{\alpha}+\omega^{2}\psi_{\alpha}= V_{_T}\, \psi_{\alpha},
      \end{equation}
    where the effective potential $V_{_T}$ is given by
      \begin{equation}\label{equation210}
       V_{_T} = q^{2}f+e^{\mathscr{A}/2}\,\partial_{r_{\ast}}^{2}e^{-\mathscr{A}/2}.
      \end{equation}

\subsection{The asymptotic wave functions}
\label{wavefunctions}

Let us investigate now the asymptotic behavior of the solutions of the
Schr\"odinger like equations \eqref{equation207} and \eqref{equation209}.
At the horizon the potentials vanish and one has free particle solutions.
Then, when $\zeta \to \zeta_h $ we have the approximate solutions  
    \begin{equation}\label{equation403}
     \psi_j \sim \mathfrak{C}_j\, e^{-i\omega r_*}+\mathfrak{D}_j\, e^{+i\omega r_*}
     \qquad\quad(j=x,y,z),
    \end{equation}
\noindent that corresponds to the superposition of incoming and outgoing waves.  
Near the horizon, the incoming $\psi^{\scriptscriptstyle{(-)}}_{j}$ and outgoing
$\psi^{\scriptscriptstyle{(+)}}_{j}$ solutions for the Schr\"odinger equations
\eqref{equation207} and \eqref{equation209} have the form 
    \begin{equation}
    \psi^{\scriptscriptstyle{(\pm)}}_{j}(\zeta)
    =e^{\pm i\omega r_{*}}\left[1+a^{\scriptscriptstyle{(\pm)}}_{1j}\left(1-\frac{\zeta}{\zeta_h}\right)
    +a^{\scriptscriptstyle{(\pm)}}_{2j}\left(1-\frac{\zeta}{\zeta_h}\right)^2+\cdots\right],
    \end{equation}
\noindent where
    \begin{equation}\label{equation409}
    \begin{split}
     a^{\scriptscriptstyle{(\pm)}}_{1j} & =\frac{1}{4\pm 2i\omega\zeta_{h}}
     \left[2\left(1+4\frac{q^2}{\omega^2}\,\delta_{jz}\right)
     +\left(q^2+4c\right)\zeta_{h}^2\right],\\
 a^{\scriptscriptstyle{(\pm)}}_{2j} & =-\frac{1}{16\pm 4i\omega\zeta_{h}}
 \left[1-12a^{\scriptscriptstyle{(\pm)}}_{1j}+16c\zeta_{h}^2
 \left(1-\frac{1}{4}c\zeta_{h}^2-\frac{q^2}{\omega^2}\,\delta_{jz}\right)+16\frac{q^2}{\omega^2}
 \left(4-5\frac{q^2}{\omega^2}\right)\delta_{jz}\right],
     \end{split}
     \end{equation}
and $\delta_{jz}$ is the Kronecker delta.

On the other hand, near the boundary the Schr\"odinger equations \eqref{equation207}
and \eqref{equation209} can be solved in terms of {\textit{normalizable}}
$\psi^{\scriptscriptstyle{(1)}}_j$ and {\textit{non-normalizable}} $\psi^{\scriptscriptstyle{(2)}}_j$
solutions, which have the asymptotic forms 
    \begin{equation}\label{equation4035}
     \hspace{-3.2cm}\psi^{\scriptscriptstyle{(1)}}_j=\left(\frac{\zeta}{\zeta_h}\right)^{3/2}\left[1+
    b_{2j}\left(\frac{\zeta}{\zeta_h}\right)^{2}+
    b_{4j}\left(\frac{\zeta}{\zeta_h}\right)^4+\cdots\right],\hspace{1.7cm}
    \end{equation}
    \begin{equation}\label{equation4036}
     \psi^{\scriptscriptstyle{(2)}}_j=\left(\frac{\zeta}{\zeta_h}\right)^{-1/2}
     \left[1+c_{2j}\left(\frac{\zeta}{\zeta_h}\right)^{2}+
    c_{4j}\left(\frac{\zeta}{\zeta_h}\right)^4
    +\cdots\right]+2d_j\,\psi^{\scriptscriptstyle{(1)}}_j\,\ln\left(\frac{\zeta}{\zeta_h}\right),
    \end{equation}
\noindent where the coefficients $c_{2j}$ are arbitrary and, for convenience, we can choose $c_{2j}=0$.
The other coefficients appearing in eqs. \eqref{equation4035} and \eqref{equation4036} are given by
%
    \begin{equation}\label{equation4037}
     \begin{split}
      b_{2j}&=-\frac{1}{8}(\omega^2-q^2)\zeta_h^2,   \quad \quad
    b_{4j}=\frac{1}{192} \left[64+8 c^2 \zeta_h^4+\left(\omega^2-q^2\right)^2 \zeta_h^4-
    \frac{32 q^2}{\omega^2-q^2}\,\delta_{jz}\right], \\
      \hspace{-6cm}c_{4j}&=\frac{1}{64} \left[8 c^2 \zeta_h^4-3 \left(\omega^2-q^2\right)^2 \zeta_h^4
      -\frac{32 q^2}{\omega^2-q^2}\,\delta_{jz}\right], \quad\quad 
      d_j=-\frac{1}{4}(\omega^2-q^2)\zeta_h^2\hspace{0.2cm}.
     \end{split}
    \end{equation}

The ingoing and outgoing wave functions can be represented using
the normalizable and non-normalizable solutions as basis,    
    \begin{equation}\label{equation4057}
      \psi_{j}^{\scriptscriptstyle{(\pm)}}=\mathfrak{A}^{\scriptscriptstyle{(\pm)}}_{j}
      \psi^{\scriptscriptstyle{(2)}}_{j}+
      \mathfrak{B}^{\scriptscriptstyle{(\pm)}}_{j}\psi^{\scriptscriptstyle{(1)}}_{j}.
    \end{equation}
\noindent The inverse relations are given by 
\begin{equation}
\psi^{\scriptscriptstyle{(1)}}_{j}=\mathfrak{C}^{\scriptscriptstyle{(1)}}_{j}
\psi^{\scriptscriptstyle{(-)}}_{j}+\mathfrak{D}^{\scriptscriptstyle{(1)}}_{j}
\psi^{\scriptscriptstyle{(+)}}_{j}\,,
\qquad\psi^{\scriptscriptstyle{(2)}}_{j}=\mathfrak{C}^{\scriptscriptstyle{(2)}}_{j}
\psi^{\scriptscriptstyle{(-)}}_{j}+\mathfrak{D}^{\scriptscriptstyle{(2)}}_{j}
\psi^{\scriptscriptstyle{(+)}}_{j}\,,
\label{inverserelation}
\end{equation}
and, as a consequence, the connection coefficients are related by 
    \begin{equation}\label{coeficientesconexao}
    \left(\begin{array}{ll}
    \mathfrak{A}^{\scriptscriptstyle{(-)}}_{j} & \mathfrak{B}^{\scriptscriptstyle{(-)}}_{j}\\
    \mathfrak{A}^{\scriptscriptstyle{(+)}}_{j} & \mathfrak{B}^{\scriptscriptstyle{(+)}}_{j}
    \end{array}\right)=
    \left(\begin{array}{ll}
    \mathfrak{C}^{\scriptscriptstyle{(2)}}_{j} & \mathfrak{D}^{\scriptscriptstyle{(2)}}_{j}\\
    \mathfrak{C}^{\scriptscriptstyle{(1)}}_{j} & \mathfrak{D}^{\scriptscriptstyle{(1)}}_{j}
    \end{array}\right)^{-1}\,.
    \end{equation}
\noindent We will need to compute numerically these coefficients
in order to obtain the vector meson spectral functions in the next section. 

The asymptotic behavior of the electric field components can be determined from the
solutions of the Schr\"odinger equations \eqref{equation4035} and \eqref{equation4036}
using the Bogoliubov transformations
\begin{equation}
E_z=e^{\mathscr{B}/2}\,\psi_{z},\qquad\quad E_\alpha=e^{\mathscr{A}/2}\,\psi_{\alpha}\qquad (\alpha=x,\,y). 
\end{equation}
Near the horizon the electric field components 
have the same decomposition in terms of incoming and outgoing waves as in eq.
(\ref{equation403}). Classically black branes only absorb radiation.
So we just have incoming electric field components at the horizon.
The solutions near the boundary for the electric field can be written in terms
of $\psi^{\scriptscriptstyle{(1)}}_j$ and $\psi^{\scriptscriptstyle{(2)}}_j$ as
      \begin{equation}
      E_{j}^{\scriptscriptstyle{(-)}}=e^{c\,\zeta^{2}/2}\,
      \left(\omega^2-q^2f\right)^{\frac{1}{2}\delta_{jz}}
      \left[\mathfrak{A}_{j}^{\scriptscriptstyle{(-)}}\psi^{\scriptscriptstyle{(2)}}_{j}+
      \mathfrak{B}_{j}^{\scriptscriptstyle{(-)}}
      \psi^{\scriptscriptstyle{(1)}}_{j}\right]\left(\frac{\zeta}{R}\right)^{1/2}\qquad(j=x,y,z),
      \label{equation31}
      \end{equation}
\noindent where the superscript $(-)$ indicates that $E_{j}^{\scriptscriptstyle{(-)}}$ satisfies 
the incoming wave condition at the horizon. These results will be used in the section \ref{Greenfunction}, 
to find the correlation functions in the dual field theory.


\subsection{An analysis of the effective potentials}
\label{potentials}

Now we will study the longitudinal and transverse potentials that appear respectively 
in the Schr\"odinger equations (\ref{equation207}) and (\ref{equation209}) in a similar way as it was done
in refs. \cite{Gibbons:2002pq,Kodama:2003jz,Morgan:2009pn}.
In terms of $\zeta$, the potentials $V_{_T}$ and $V_{_L}$ can be written explicitly as
\begin{equation}
V_{_T}=\frac{f}{\zeta^2}\left[q^2\zeta^2+\frac{3}{4}+\left(\frac{5}{4}+4c\zeta^2\right)
\left(\frac{\zeta}{\zeta_h}\right)^4+c^2\zeta^4 f\right],
\label{Vtransversal}
\end{equation}
\begin{equation}
V_{_L}=V_{_T}+\frac{4 q^2 f}{\zeta^2(\omega^2-q^2 f)^2}\left[q^2 f+c\zeta^2f(\omega^2-q^2 f)-\omega^2
\left(1-3\frac{\zeta^4}{\zeta_{h}^4}\right)\right]\left(\frac{\zeta}{\zeta_h}\right)^4 .
\label{Vlongitudinal}
\end{equation}

Let us consider some important particular cases for the temperature, for the parameter $c$
and for the wave number $q$. At zero temperature these expressions reproduce the original
soft-wall potential of \cite{Karch:2006pv},  
%
    \begin{equation}\label{potencialtzero}
     V_{_L} (\zeta)\big|_{T=0}=V_{_T} (\zeta)\big|_{T=0}=\frac{1}{\zeta^2}
     \left[q^2\zeta^2+\frac{3}{4}+c^2 \zeta^4\right]\,.
    \end{equation}
\noindent On the other hand, in the limit of vanishing infrared cut off $c$,
the potentials become  
\begin{equation}
V_{_{_T}} (\zeta)\big|_{c=0}=\frac{f}{\zeta^2}\left[q^2\zeta^2+\frac{3}{4}+\frac{5}{4}
\left(\frac{\zeta}{\zeta_h}\right)^4\right],
\end{equation}
\begin{equation}
V_{_L} (\zeta)\big|_{c=0}=V_{_T} (\zeta)\big|_{c=0}+\frac{4 q^2 f}{(\omega^2-q^2 f)^2}
\left[q^2 f-\omega^2\left(1-3\frac{\zeta^4}{\zeta_{h}^4}\right)\right]\left(\frac{\zeta}{\zeta_h}\right)^4 .
    \end{equation}
which are the potentials governing the evolution of electromagnetic perturbations in
the spacetime of a planar $\mbox{AdS}_5$ black hole.

Another interesting case corresponds to vanishing wave number $q$, for which
the potential (\ref{Vtransversal}) coincides with the longitudinal case
given in eq. (\ref{Vlongitudinal}). Then we define:

\begin{equation}
V (\zeta )\equiv V_{_T} (\zeta)\big|_{q=0}=V_{_L} (\zeta)\big|_{q=0}\,,
\end{equation}
so that
    \begin{equation}\label{potqzero}
    V(\zeta)=\frac{f}{\zeta^2}\left[\frac{3}{4}+\left(\frac{5}{4}+4c\zeta^2\right)
\left(\frac{\zeta}{\zeta_h}\right)^4+c^2\zeta^4 f\right].
    \end{equation}
\noindent Note that in this particular case the potential does not depend on the energy $\omega$.
Furthermore, this case can be interpreted as the potential associated with quasiparticles 
at rest in the dual gauge theory. 

Now we want to analyse the form of the potential of eq. (\ref{potqzero}) 
for different temperature regimes. For this purpose it is convenient to work with dimensionless 
quantities. The potentials that appear in the Schr\"odinger like equations (\ref{equation207}) and 
(\ref{equation209}) have dimension of energy squared. So we can write dimensionless equations
by dividing $V$ by  $c$. This leads us naturally to the dimensionless coordinates $ \sqrt{c}\, \zeta $ and
$ \sqrt{c}\, r_{*}\,$.
Then the corresponding dimensionless temperature is $\widetilde{T}=\pi T / \sqrt{c}\, $.   
The high temperature regime corresponding to $\widetilde{T}^2 > 1$ is characterized by the 
potential shape of an infinite barrier as shown in Figure \ref{potencialhigh}. 
One also observes that the potential increases with the temperature and with the tortoise coordinate. 
As discussed in the next section, there are no quasiparticle states in the dual theory
in this regime.

\FIGURE{
\centering
    \includegraphics[width=7cm,angle=-90]{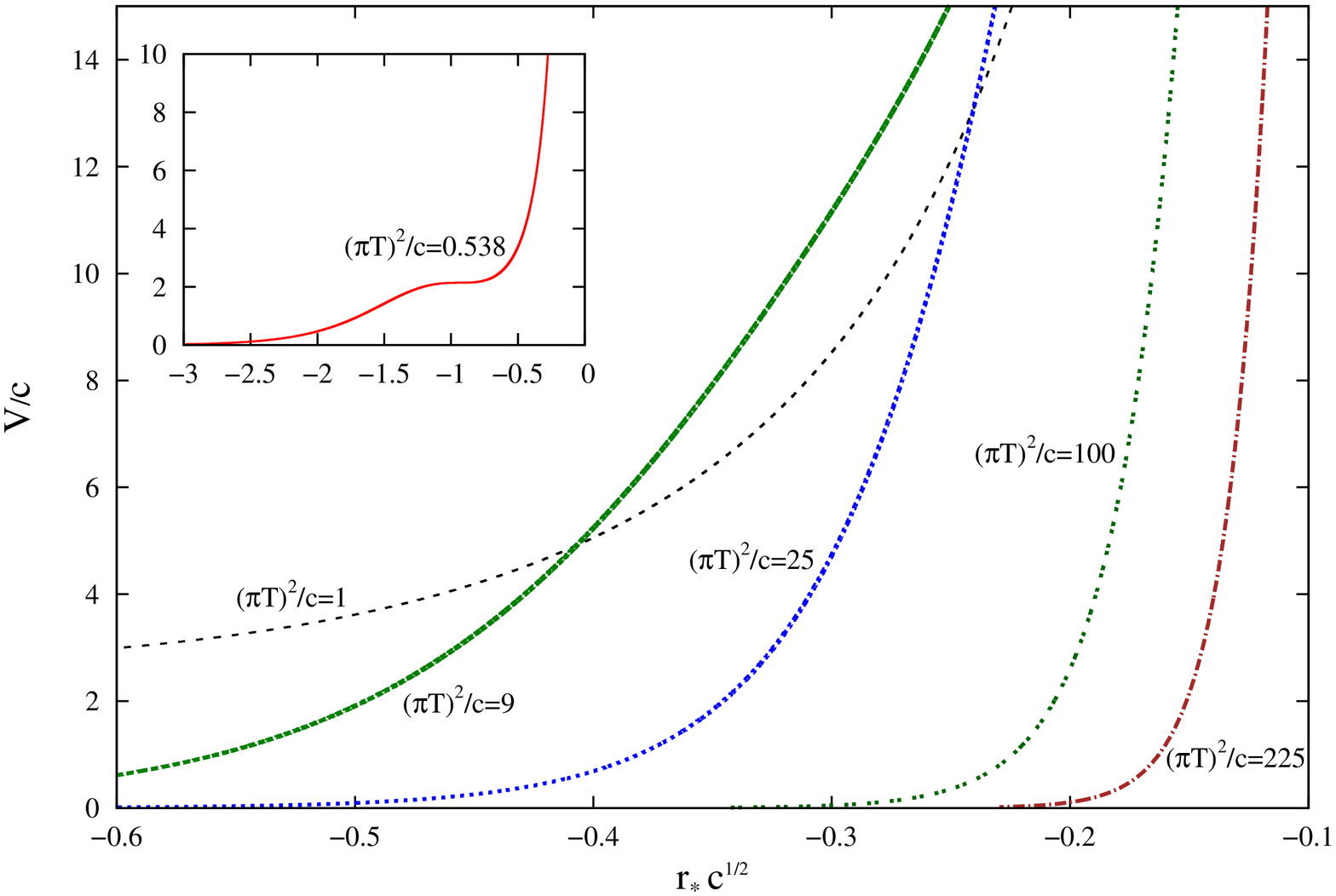}
\centering\caption{Potential at zero wave number for high temperatures.
    We also show the potential at the critical value $\widetilde{T}^2_c = 0.538$ in the detail.}
\label{potencialhigh}
}

For smaller values of $\widetilde{T}$ the behavior of the potential changes.
Below the critical value $\widetilde{T}^2_c = 0.538$ the potential presents a well
as illustrated in figure \ref{potenciallow}. The depth of the well increases as the
temperature decrease.  One can also see that there is an infinite barrier
localized near the boundary ($r_*=0$). In this regime there are quasiparticle states
in the dual theory, corresponding to vector mesons at finite temperature (see section
\ref{Greenfunction} below). The probability of vector meson formation increases
when the temperature decreases.

To analyse the low temperature regime it is convenient to write $V(\zeta)$ as a power series expansion
in $\widetilde{T}$. Changing $\zeta_h$ by $T=1/\pi\zeta_h$ in \eqref{potqzero},
the potential takes the form
    \begin{equation}\label{potencialexpandido}
      V=\frac{3}{4 \zeta^2}+c^2 \zeta^2+\left(\frac{1}{2}+4c\zeta^2-2c^2\zeta^4\right)\pi^4 \zeta^2 T^4
      -\left(\frac{5}{4}+4c\zeta^2-c^2\zeta^4\right)\pi ^8 \zeta^6 T^8.
    \end{equation}
We can also express this potential in terms of the tortoise coordinate (\ref{equation206}) which,
near the boundary, can be written as \cite{Miranda:2009uw}
    \begin{equation}\label{tartarugafronteira}
     \zeta=-r_*\left[1-\frac{r_* ^4}{5 \zeta_h ^4}\right]\,,
    \end{equation}
so that keeping only the leading order term on the temperature we have    
    \begin{equation}
     V=\frac{3}{4 r_*^2}+c^2 r_*^2+\frac{4}{5}  \left(1+5c r_*^2 -3 c^2 r_*^4\right)\pi ^4 r_*^2 T^4.
    \end{equation}
One notes that this potential is a sum of terms that correspond to: an infinite barrier 
localized at $r_* \rightarrow 0$, a harmonic oscillator like term $r_*^2$, and the temperature  
contributions. The superposition of these terms gives the potential with the form shown in figure 
\ref{potenciallow}. 
 
\FIGURE{
\centering\includegraphics[width=7cm,angle=-90]{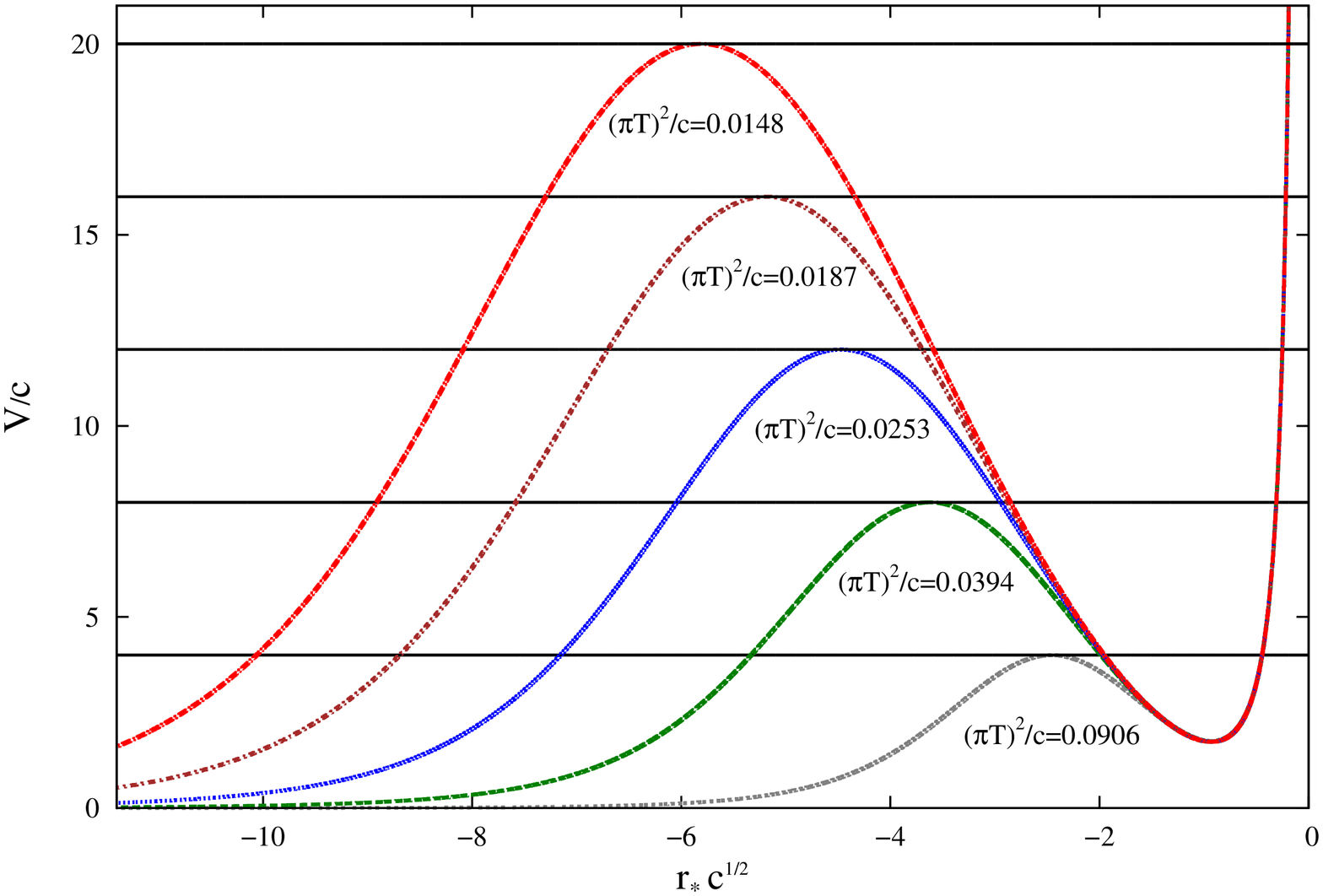}    
\centering\caption{Potential at zero wave number for low temperatures.}
\label{potenciallow}
}


\section{The retarded Green's functions}
\label{Greenfunction}

\subsection{Definitions and analytical results}

In a quantum field theory at finite temperature, the  Fourier-transformed retarded
Green's functions of conserved symmetry currents are defined as
      \begin{equation}\label{correl}
      G^{\scriptscriptstyle{R}}_{\mu \nu}(k) = -i \int d^4 x e^{- ik\cdot x} \theta (t) \, 
\langle \left [ J_\mu (x)  , J_\nu (0)   \right ]\rangle\,.
      \end{equation}
The functions $G^{\scriptscriptstyle{R}}_{\mu \nu}(k)$ can be separated into transverse
and longitudinal parts \cite{Kovtun:2005ev},
      \begin{equation}\label{correlatores}
      G^{\scriptscriptstyle{R}}_{\mu\nu}(k)=P_{\mu\nu}^{T}\,\Pi^{T}(k_{0},\textbf{k}^2)+
      P_{\mu\nu}^{L}\,\Pi^{L}(k_{0},\textbf{k}^2)\,,
      \end{equation}
where $\Pi^{T}(k_{0},\textbf{k}^2) $ and $\Pi^{L}(k_{0},\textbf{k}^2)$ are
independent scalar functions and the projectors are given by 
\begin{equation}
\begin{aligned}[2]
& P_{\mu\nu}=\eta_{\mu\nu}-\frac{k_{\mu}k_{\nu}}{k^2} \,,\qquad  \qquad 
& P^{L}_{\mu\nu}=P_{\mu\nu}-P_{\mu\nu}^{T}\, ,\\
&P^{T}_{00}=0 \,, \;\;\,\quad\quad P^{T}_{0i}=0,
& P^{T}_{ij}=\delta_{ij}-\frac{k_{i}k_{j}}{\bf{k}^2}\,.
\end{aligned}
\end{equation}
Note that these projectors satisfy $k^{\mu}P_{\mu\nu}^{T}=k^{\mu}P_{\mu\nu}^{L}=0$,
implying current conservation. 

We can consider, without loss of generality, that the wave vector has the form 
$k_{\mu}=(-\omega,0,0,q)$, corresponding, as considered in the previous section,
to the propagation in the $z$ direction. Then the non vanishing components of
the current-current correlation function are
      \begin{equation}\label{correlatortransversal}
      G^{\scriptscriptstyle{R}}_{xx}(k)=G^{\scriptscriptstyle{R}}_{yy}(k)=\Pi^{T}(\omega,q)\,,
      \end{equation}
     \begin{equation}\label{correlatorlongitudinal}
      G^{\scriptscriptstyle{R}}_{tt}(k)=\frac{q^2}{\omega^2-q^2}\Pi^{L}(\omega,q),\quad
      G^{\scriptscriptstyle{R}}_{tz}(k)=-\frac{q\omega}{\omega^2-q^2}\Pi^{L}(\omega,q),\quad
      G^{\scriptscriptstyle{R}}_{zz}(k)=\frac{\omega^2}{\omega^2-q^2}\Pi^{L}(\omega,q).
      \end{equation}
On the other side, these correlation functions of the four dimensional field theory can be  
obtained from the dual bulk fields. As an illustration, for scalar fields, following the prescriptions
of Ref. \cite{Son:2002sd} one writes the on shell action in the form 
    \begin{equation}\label{presc}
     S=\int\frac{d^4k}{(2\pi)^4}\phi_0(-k)\mathcal F(k,\zeta)\phi_0(k)\bigg|_{\zeta=\zeta_B}^{\zeta=\zeta_h},
    \end{equation}
where $\phi_0(k)$ is the boundary value of the field. The ``flux factor'' $\mathcal F$ is
    \begin{equation}\label{funcaofronteira}
     \mathcal F(\zeta,k)=K\sqrt{-g}\,g^{{{\zeta}}{{\zeta}}}f_{-k}(\zeta)\partial_{{\zeta}}f_{k}(\zeta),
    \end{equation}
where $f_k(\zeta)=\phi(\zeta,k)/\phi_{0}(k)$ satisfies an incoming-wave condition
at the horizon and, by definition, is normalized to unity at
the boundary. According to this prescription, the retarded Green's function is
related to the flux factor by
    \begin{equation}\label{funcaogreen}
     G^R(k)\equiv -2\mathcal F(k,\zeta)\Big|_{\zeta_B},
    \end{equation}
where $\zeta_B$ is the value of the coordinate $\zeta$ at the boundary, which in 
the present case is zero.

In our case, for the vector field, we start writing the action (\ref{action}) in terms
of the Fourier transforms of the components of the vector fields in the radial
gauge $A_{\zeta}=0$,
      \begin{equation*}\label{equation38}
      S=\frac{R}{2g_{5}^2}\int \frac{d\omega dq}{(2\pi)^2}\
      \Bigg[\frac{e^{-\Phi}}{\zeta}\Big\{A_{t}(\zeta,-k)\partial_{{\zeta}}A_{t}(\zeta,k)-
      f\mathbf{A}(\zeta,-k)\cdot\partial_{{\zeta}}\mathbf{A}(\zeta,k)\Big\}
      \Bigg]_{0}^{\zeta_{h}}\,.
      \end{equation*}
Expressing this action in terms of the electric field components we have   
     \begin{equation}\label{equation39} 
      S=-\frac{N_{c}^2}{32\pi^2}\int \frac{d\omega dq}{(2\pi)^2}
      \Bigg[\frac{e^{-\Phi}}{\zeta}\frac{f}{{\omega}^2}\sum_{j}
      \left(1-\frac{q^2}{\omega^2}f\right)^{-\delta_{jz}}
      E_{j}(\zeta,-k)\,\partial_{{\zeta}}\,E_{j}(\zeta,k)\Bigg]_{0}^{\zeta_{h}}\,,
      \end{equation}
where above we have used the relation $R/2g_{5}^{2}=N_{c}^{2}/32\pi^2$.
The bulk electric field components can be written in terms of their boundary values
$E_{j}^0$ as
      \begin{equation}\label{relationEj}
      E_{j}^{\scriptscriptstyle{(-)}}(\zeta,k)={\cal E}_{j}(\zeta,k)\,E^{0}_{j}(k),\qquad\quad(j=x,y,z),
      \end{equation}
where the functions ${\cal E}_{j}(\zeta,k)$ are defined so that ${\cal E}_{j}(0,k)=1$
and the superscript $(-)$ indicates that $E_{j}^{\scriptscriptstyle{(-)}}(\zeta,k)$
satisfies the ingoing wave condition at the horizon, as required by the
Lorentzian Son-Starinets prescription \cite{Son:2002sd}.
Substituting \eqref{relationEj} in the action (\ref{equation39})
one finds
     \begin{equation}\label{equation40} 
      S=-\frac{N_{c}^2}{32\pi^2}\int \frac{d\omega dq}{(2\pi)^2}
      \Bigg[\frac{e^{-\Phi}}{\zeta}\frac{f}{{\omega}^2}\sum_{j}
      \left(1-\frac{q^2}{\omega^2}f\right)^{-\delta_{jz}}
      E^{0}_{j}(-k){\cal E}_{j}(\zeta,-k)\,\partial_{{\zeta}}\,{\cal E}_{j}(\zeta,k)\,
      E^{0}_{j}(k)\Bigg]_{0}^{\zeta_{h}}\,.
      \end{equation}
In terms of the boundary values of the potential this action reads
      \begin{equation}
      \begin{split}
      & S=\frac{N_{c}^2}{32\pi^2}\int \frac{d\omega dq}{(2\pi)^2}
      \Bigg[\frac{e^{-\Phi}}{\zeta}f\left(\frac{\omega^2}{\omega^{2}-q^{2}f}\Big\{A^{0}_{z}(-k)\,      
      A^{0}_{z}(k)+\frac{q}{\omega} A^{0}_{z}(-k)A^{0}_{t}(k)+\frac{q}{\omega}A^{0}_{t}(-k)
      A^{0}_{z}(k)\right. \\
      &+\frac{q^2}{\omega^2} A^{0}_{t}(-k)A^{0}_{t}(k)\Big\}
      {\mathcal E}_{z}(\zeta,-k)\partial_{{\zeta}}{\mathcal E}_{z}(\zeta,k)
      +\sum_{\alpha}A^{0}_{\alpha}(-k)A^{0}_{\alpha}(k){\mathcal E}_{\alpha}(\zeta,-k)
      \partial_{{\zeta}}{\mathcal E}_{\alpha}(\zeta,k)
      \bigg)\Bigg]_{0}^{\zeta_{h}}\,.
      \end{split}
      \end{equation}
Using the Son-Starinets (\ref{funcaogreen}) prescription for the vector field case
we find the current current correlators $G_{\mu\nu}^{\scriptscriptstyle{R}}(k)
\equiv-2\,\eta_{\mu\lambda}\,\eta_{\nu\kappa}\,\mathcal{F}^{\lambda\kappa}(k,\zeta)\big|_{\zeta=0}$,
so that
\begin{equation}\label{equation41}
 \begin{split}
   & \frac{G^{\scriptscriptstyle{R}}_{tt}}{q^2}=-\frac{G^{\scriptscriptstyle{R}}_{tz}}{\omega q}=
     -\frac{G^{\scriptscriptstyle{R}}_{zt}}{\omega q}=
     \frac{G^{\scriptscriptstyle{R}}_{zz}}{\omega^2}=-\frac{N_{c}^2}{16\pi^2(\omega^2-q^2)}
     \lim_{\zeta\rightarrow 0}\frac{1}{\zeta}\,\partial_{{\zeta}}\,{\mathcal E}_{z}(\zeta,k),\\
     & G^{\scriptscriptstyle{R}}_{xx}=-\frac{N_{c}^2}{16\pi^2}
     \lim_{\zeta\rightarrow 0}\frac{1}{\zeta}\,\partial_{{\zeta}}\,{\mathcal E}_{x}(\zeta,k),
     \quad G^{\scriptscriptstyle{R}}_{yy}=-\frac{N_{c}^2}{16\pi^2}
     \lim_{\zeta\rightarrow 0}\frac{1}{\zeta}\,\partial_{{\zeta}}\,{\mathcal E}_{y}(\zeta,k).
\end{split}
\end{equation} 
Considering the form of the electric fields near the
boundary given in eq. (\ref{equation31}) one then finds 
 \begin{equation}
      {\cal{E}}_{j}=e^{c\,\zeta^{2}/2}\,
      \left(\frac{\omega^2-q^2f}{\omega^2-q^2}\right)^{\frac{1}{2}\delta_{jz}}
      \left[\psi^{\scriptscriptstyle{(2)}}_{j}+
      \frac{\mathfrak{B}_{j}^{\scriptscriptstyle{(-)}}}{\mathfrak{A}_{j}^{\scriptscriptstyle{(-)}}}
      \psi^{\scriptscriptstyle{(1)}}_{j}\right]\left(\frac{\zeta}{\zeta_h}\right)^{1/2}.
      \label{equation42}
      \end{equation}
Substituting \eqref{equation42} in \eqref{equation41} and comparing the resulting expressions
with \eqref{correlatortransversal} and \eqref{correlatorlongitudinal}
we get the longitudinal and transversal parts of the current current correlators 
    \begin{equation}\label{equation43a}
      \Pi^{L}(\omega,q)=-\frac{N_{c}^{2}T^2}{8}\left\{\frac{1}{2}c\zeta_{h}^{2}+
      d\left[1+2\ln\left(\frac{\epsilon}{\zeta_h}\right)\right]+
      \frac{\mathfrak{B}^{\scriptscriptstyle{(-)}}_{z}}{\mathfrak{A}^{\scriptscriptstyle{(-)}}_{z}}
      \right\},
    \end{equation}
    \begin{equation}\label{equation43b}
      \Pi^{T}(\omega,q)=-\frac{N_{c}^{2}T^2}{8}\left\{\frac{1}{2}c\zeta_{h}^{2}+
      d\left[1+2\ln\left(\frac{\epsilon}{\zeta_h}\right)\right]+
      \frac{\mathfrak{B}^{\scriptscriptstyle{(-)}}_{x}}{\mathfrak{A}^{\scriptscriptstyle{(-)}}_{x}}
      \right\},
    \end{equation}
where $d\equiv d_j=-(\omega^2-q^2)\zeta_{h}^{2}/4$ and $\epsilon$ is an ultraviolet
regulator. The coefficient $d$ is an analytic function of $\omega$ and $q$, and so
the terms containing $d$ in \eqref{equation43a} and \eqref{equation43b} are contact
terms that can be removed from the Green's functions by means of a holographic
renormalization \cite{Bianchi:2001kw}, i.e., the addition of boundary counterterms
to the action \eqref{action}.

\subsection{Spectral functions at finite temperature}

\subsubsection{General procedure}

The imaginary parts of the retarded Green's functions give us the 
spectral functions. Particularly, in the case of $G^{\scriptscriptstyle{R}}_{zz}$
and $G^{\scriptscriptstyle{R}}_{xx}$, the corresponding spectral functions are given by
    \begin{equation}\label{R1}
     \mathfrak{R}_{zz}(\omega,q)\equiv -2\,\text{Im}G_{zz}^{\scriptscriptstyle{R}}(\omega,q)=
     \frac{N_{c}^{2}T^{2}}{4}\left(\frac{\omega^2}{\omega^2-q^2}\right)\text{Im}
     \left(\frac{\mathfrak{B}^{\scriptscriptstyle{(-)}}_z}{\mathfrak{A}^{\scriptscriptstyle{(-)}}_z}\right),
    \end{equation}
    \begin{equation}\label{R2}
    \mathfrak{R}_{xx}(\omega,q)\equiv -2\,\text{Im}G_{xx}^{\scriptscriptstyle{R}}(\omega,q)=
     \frac{N_{c}^{2}T^{2}}{4}\text{Im}
     \left(\frac{\mathfrak{B}^{\scriptscriptstyle{(-)}}_x}{\mathfrak{A}^{\scriptscriptstyle{(-)}}_x}\right).
    \end{equation}

Now we have to find out the above quantities in terms of the solutions of the equations of
motion (\ref{equation207}) and (\ref{equation209}). So we will need the power series
expressions for the normalizable and non-normalizable solutions of the Schr\"odinger like
equations \eqref{equation207} and \eqref{equation209}, that means, eqs. (\ref{equation4035})
and (\ref{equation4036}) and also the corresponding derivatives with respect to $\zeta$
that read:
    \begin{equation}
     \partial_{\zeta}\psi^{\scriptscriptstyle{(1)}}_{j}(\zeta_{nb})=
     \frac{1}{2}\Big(\frac{\zeta_{nb}}{\zeta_h}\Big)^{1/2}\left[3+
     7\,b_{2j}\Big(\frac{\zeta_{nb}}{\zeta_h}\Big)^2+11\,b_{4j}
     \Big(\frac{\zeta_{nb}}{\zeta_h}\Big)^4\right],
     \label{E1}
    \end{equation}
     \begin{equation}
 \partial_{\zeta}\psi^{\scriptscriptstyle{(2)}}_{j}(\zeta_{nb})=\frac{1}{2}
 \Big(\frac{\zeta_{nb}}{\zeta_h}\Big)^{-3/2}\left[-1+7\,c_{4j}
 \Big(\frac{\zeta_{nb}}{\zeta_h}\Big)^4\right]+2d_{j}\left[\partial_{\zeta}
 \psi^{\scriptscriptstyle{(1)}}_{j}(\zeta_{nb})
 \ln\Big(\frac{\zeta_{nb}}{\zeta_h}\Big)
 +\frac{\zeta_h}{\zeta_{nb}}\psi^{\scriptscriptstyle{(1)}}_{j}(\zeta_{nb})\right],
 \label{E2}
       \end{equation}
where $\zeta_{nb} $ is a value for $\zeta$ near the boundary and the coefficients $b_{2j}$, $b_{4j}$,
$c_{4j}$ and $d_{j}$ were determined in eq. (\ref{equation4037}).

We will perform a numerical integration of the Schr\"odinger equations (\ref{equation207})
and (\ref{equation209}), using $\psi^{\scriptscriptstyle{(1)}}_{j}(\zeta_{nb})$,
$\psi^{\scriptscriptstyle{(2)}}_{j}(\zeta_{nb})$ and their derivatives as ``initial'' conditions,
and integrating from a point near the boundary $\zeta_{nb}$ to a point near the horizon $\zeta_{nh}$. 
Then we express the functions $\psi^{\scriptscriptstyle{(1)}}_{j}$ and
$\psi^{\scriptscriptstyle{(2)}}_{j}$ in terms of the ingoing
and outgoing solutions using eq. (\ref{inverserelation}) at the point $\zeta_{nh}$ in the form
    \begin{equation}\label{matrizes1}
    \left(\begin{array}{cc}
    \psi^{\scriptscriptstyle{(a)}}_{j}(\zeta_{nh})\\
    \partial_{\zeta}\psi^{\scriptscriptstyle{(a)}}_{j}(\zeta_{nh})
    \end{array}\right)=
    \left(\begin{array}{cc}
    \psi_{j}^{\scriptscriptstyle{(-)}}(\zeta_{nh}) \quad& \quad \psi_{j}^{\scriptscriptstyle{(+)}}(\zeta_{nh})\\
    \partial_{\zeta}\psi_{j}^{\scriptscriptstyle{(-)}}(\zeta_{nh})\quad &
    \quad \partial_{\zeta }\psi_{j}^{\scriptscriptstyle{(+)}}(\zeta_{nh})
    \end{array}\right)\left(
      \begin{array}{cc}
       \mathfrak{C}^{\scriptscriptstyle{(a)}}_{j}\\
	\mathfrak{D}^{\scriptscriptstyle{(a)}}_{j}
      \end{array}\right) \qquad\quad (a=1,\,2).
    \end{equation}
The  coefficients $\mathfrak{C}^{\scriptscriptstyle{(a)}}_{j}$ and
$\mathfrak{D}^{\scriptscriptstyle{(a)}}_{j}$ can then be determined as
    \begin{equation}\label{matrizes2}\left(
      \begin{array}{cc}
       \mathfrak{C}^{\scriptscriptstyle{(a)}}_{j}\\
	\mathfrak{D}^{\scriptscriptstyle{(a)}}_{j}
      \end{array}\right)
   =
    \left(\begin{array}{cc}
    \psi_{j}^{\scriptscriptstyle{(-)}}(\zeta_{nh}) \quad & \quad \psi_{j}^{\scriptscriptstyle{(+)}}(\zeta_{nh})\\
    \partial_{\zeta}\psi_{j}^{\scriptscriptstyle{(-)}}(\zeta_{nh})\quad &
    \quad \partial_{\zeta }\psi_{j}^{\scriptscriptstyle{(+)}}(\zeta_{nh})
    \end{array}\right)^{-1}
    \left(\begin{array}{cc}
    \psi^{\scriptscriptstyle{(a)}}_{j}(\zeta_{nh})\\
    \partial_{\zeta}\psi^{\scriptscriptstyle{(a)}}_{j}(\zeta_{nh})
    \end{array}\right) \qquad\quad (a=1,\,2).
    \end{equation}
Finally, from eqs. \eqref{coeficientesconexao} and \eqref{matrizes2} we obtain the ratio
     \begin{equation}
     \frac{\mathfrak{B}^{\scriptscriptstyle{(-)}}_{j}}{\mathfrak{A}^{\scriptscriptstyle{(-)}}_{j}}=
     -\frac{\mathfrak{D}^{\scriptscriptstyle{(2)}}_{j}}{\mathfrak{D}^{\scriptscriptstyle{(1)}}_{j}}
     =-\frac{\partial_{\zeta }\psi_{j}^{\scriptscriptstyle{(-)}}\,
     \psi^{\scriptscriptstyle{(2)}}_{j}-\psi_{j}^{\scriptscriptstyle{(-)}}\,
     \partial_{\zeta}\psi^{\scriptscriptstyle{(2)}}_{j}}
     {\partial_{\zeta }\psi_{j}^{\scriptscriptstyle{(-)}}\,
     \psi^{\scriptscriptstyle{(1)}}_{j}-
     \psi_{j}^{\scriptscriptstyle{(-)}}\,
     \partial_{\zeta}\psi^{\scriptscriptstyle{(1)}}_{j}}\Bigg|_{\zeta = \zeta_{nh}}
    \end{equation}
that we need in order to determine the spectral functions, as in eqs. (\ref{R1}) and (\ref{R2}). 

\subsubsection{Numerical results}

We performed the numerical analysis of the spectral functions $\mathfrak{R}_{xx}$ and
$\mathfrak{R}_{zz}$, focusing especially on the longitudinal component of the gauge field.
We used the dimensionless quantities $\widetilde{\omega}=\omega/\sqrt{c}$,
$\widetilde{q}=q/\sqrt{c}$ and the previously defined $\widetilde{T}=\pi T/\sqrt{c}$.
We show in figure \ref{espectral1} the meson vector spectral function for a
vanishing wave number. As discussed in the subsection \ref{potentials},
the longitudinal and transverse potentials are equal for $q=0$, so that
\begin{equation}
\mathfrak{R}_{0}(\omega)\equiv\mathfrak{R}_{xx}(\omega,0)=\mathfrak{R}_{yy}(\omega,0)=\mathfrak{R}_{zz}(\omega,0).
\end{equation}
The peaks shown in figure \ref{espectral1} indicate that the corresponding retarded Green's
functions present poles. These poles are related to the frequencies of the
electromagnetic quasinormal modes of the
black brane. In the dual gauge theory, these quasinormal modes correspond to quasiparticle states 
of vector mesons. The frequencies of these quasinormal modes present real
($\widetilde{\omega}_{\scriptscriptstyle{R}}$) and imaginary
($\widetilde{\omega}_{\scriptscriptstyle{I}}$) parts. The real part is related to the mass
of the vector mesons when $q=0$, while the imaginary part to the decay rate of the
quasiparticle states formed near the  confining/deconfining transition. One notes
that when the temperature increases, the  width of the peaks increases and the mean
life $\tau =2\pi/\widetilde{\omega}_{\scriptscriptstyle{I}}$ decreases.
These results are in agreement with the analysis obtained from the effective potentials presented in the
previous section. It is interesting to mention that the peaks in figure \ref{espectral1}
are localized according to $\widetilde{\omega}^2 \approx 4n$, where $n=1,2,3\dots$, for low temperatures.
%
\FIGURE{
\centering\includegraphics[width=7cm,angle=-90]{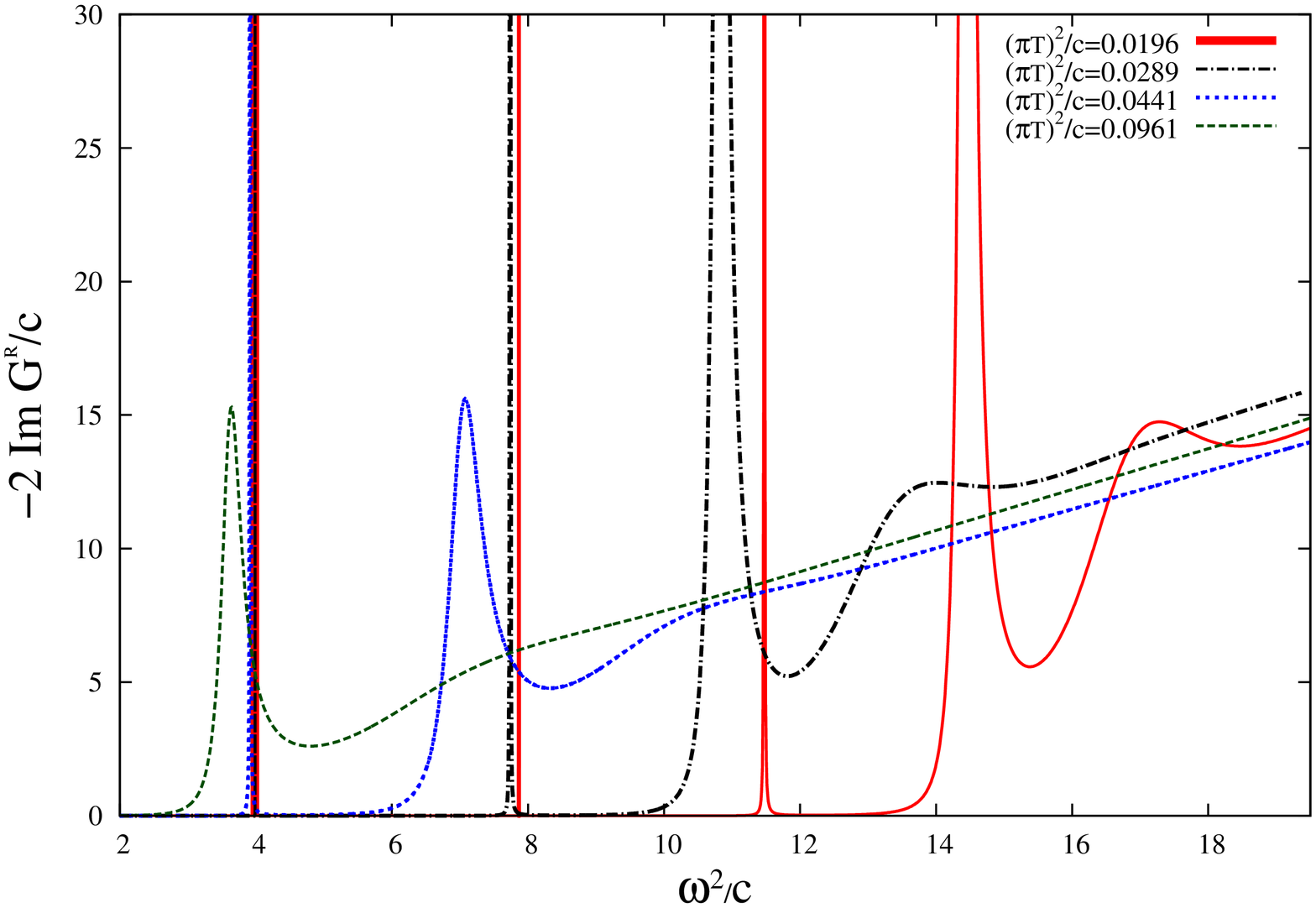}
\centering\caption{Spectral function at zero wave number as a function of the energy 
    for different values of the temperature $\widetilde{T}^2=(\pi T)^2/c$.}\label{espectral1}
}
\noindent Near the peaks, one can approximate the spectral function by a  Breit-Wigner distribution
\cite{Miranda:2009uw,Fujita:2009wc,Colangelo:2009ra}:
    \begin{equation}
    \mathfrak{R}_{0}(\omega)=\frac{\mathcal{A}\,\widetilde{\omega}^{\,\mathfrak{b}}}{(\widetilde{\omega}-
    \widetilde{\omega}_{\scriptscriptstyle{R}})^2+\widetilde{\omega}_{\scriptscriptstyle{I}}^{2}}\,,
    \end{equation}
where $\widetilde{\omega}_{\scriptscriptstyle{R}}$ is the frequency of the peak and
$\widetilde{\omega}_{\scriptscriptstyle{I}}$ is its width. The quantities
$\mathcal{A}$ and $\mathfrak{b}$ are adjustable constants that vary with the
temperature and the position of the peak in the frequency axis. For $\widetilde{T}^2=0.0441$,
$\mathcal{A}=0.0281463$ and $\mathfrak{b}=0.109005$ for the peak of lower frequency, while
$\mathcal{A}=5.65799\times 10^{-9}$ and $\mathfrak{b}=11.3886$ for the second peak
of lower frequency. One can also note from figure
\ref{espectral1} that the peaks of the spectral function depend on the temperature.
For temperatures higher than $\widetilde{T}^2_c=2.3864$ there are no peaks.
Decreasing the temperature the peaks are formed at frequencies near the zero temperature values. 

We also studied the spectral function $\mathfrak{R}_{zz}(\omega,q)$ at non vanishing wave number.
We show in figure \ref{espectral2} the results obtained for the fixed temperature $\widetilde{T}^2=0.0484$.
We note in this figure that the height of the peaks decreases as the wave number increases.
Furthermore, the width of the peaks increases with the wave number and the frequency.
This can be interpreted as meaning that quasiparticle states with momenta ($q \ne 0$)
are more unstable than those at rest ($q=0$). The localization of the peaks in terms
of the frequency also changes with the wave number. This was expected from the approximate
dispersion relation  $\widetilde{\omega}^{\,2}\approx\widetilde{\omega}_{0}^{\,2}+\widetilde{q}^{\,2}$,
where $\widetilde{\omega}_0$ is the frequency at zero wave number \cite{Fujita:2009wc}.

\FIGURE{
\centering\includegraphics[width=7cm,angle=-90]{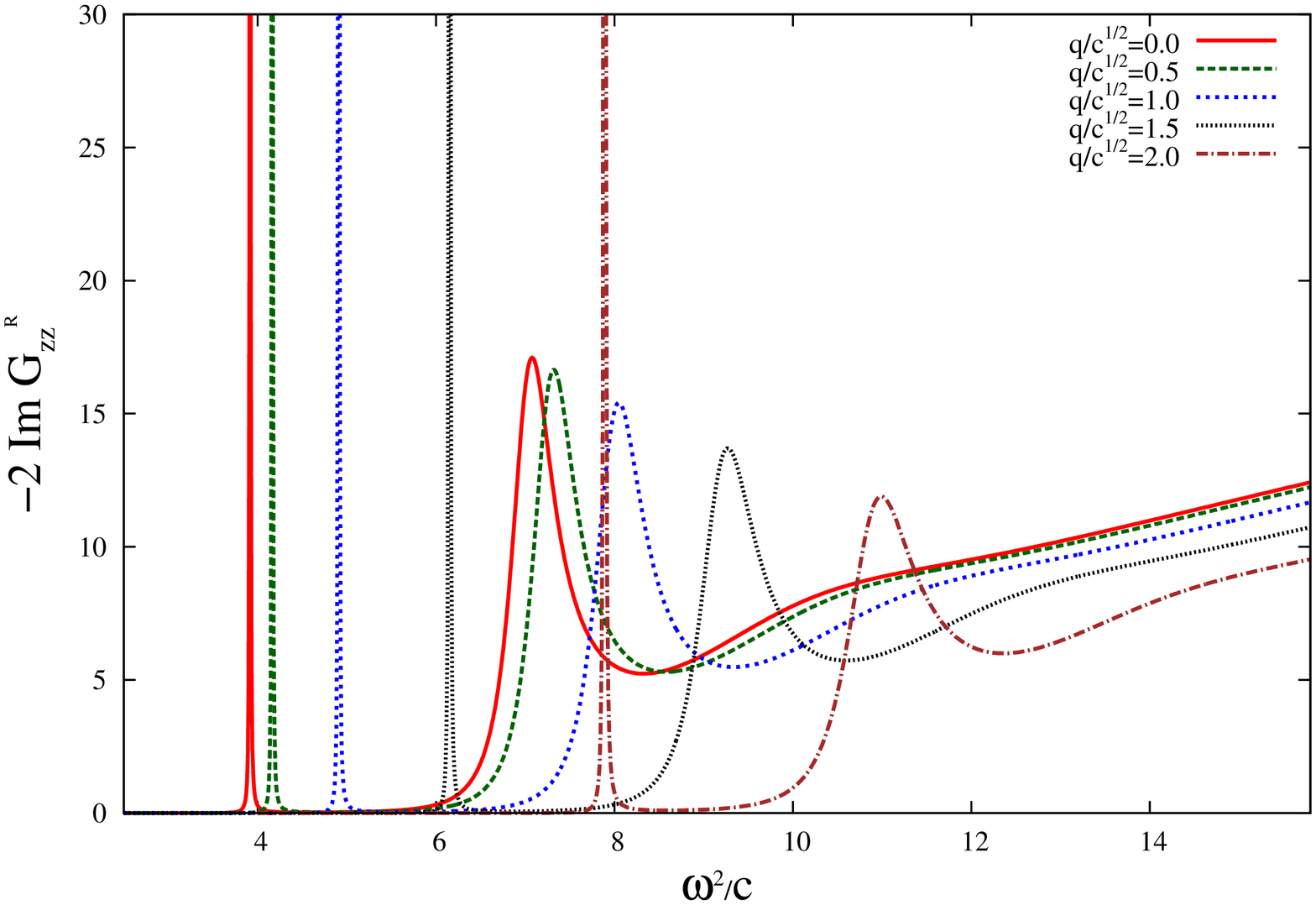}
\centering\caption{Spectral function $\mathfrak{R}_{zz}(\omega,q)$ at
     $\widetilde{T}^2=0.0484$ as a function of the energy for different values
     of the wave number.}\label{espectral2}
}

\section{Quasinormal modes} 
\label{section_modosquasenormais}

The solutions for classical field perturbations on a black hole geometry
subjected to an absorption condition at the horizon and a regularity
condition at the spacetime  boundary are called quasinormal modes (QNMs)
(see Refs. \cite{Berti:2009kk,Konoplya:2011qq} for reviews and
Refs. \cite{Maeda:2005cr,Siopsis:2004as,Miranda:2005qx,Zhang:2006bc,
Hoyos:2006gb,Miranda:2007bv,Morgan:2009vg} as examples). For an AdS black hole,
the absorption condition corresponds to a pure incoming wave at the horizon and
the regularity at the boundary is guaranteed by imposing Dirichlet conditions on the solutions. 

In this section we are going to obtain the quasinormal modes for electromagnetic perturbations 
by solving the equations of motion using first perturbative methods and then numerical methods.

\subsection{Perturbative analytical solutions }
\label{perturbativesolutions}

Here we are going to solve the equations of motion in the so called hydrodynamic regime,
that corresponds to low frequencies and low wave numbers compared to the Hawking temperature
of the black hole. We use the notation  $\mathfrak w={\omega}/{(\pi T)}$ and
$\mathfrak q={q}/{(\pi T)}$, so that in this regime:  $\mathfrak w \ll 1$ and $\mathfrak q \ll 1$.

\subsubsection{Longitudinal perturbations }
\label{solucaohidrodinamicolongitudinal}

For the longitudinal sector we write $E^{\scriptscriptstyle{(-)}}_{z}=f^{-i\mathfrak{w}/4}F_z$
and substitute in the differential equation (\ref{equation25}) obtaining 
      \begin{equation}\label{equation43}
      \begin{split}
      \partial_{\zeta}^{2}F_{z}-\bigg[1+2 {c}\zeta^2+
      \bigg(\frac{4a_z-2i\mathfrak{w}}{f}\bigg)
      \frac{\zeta^4}{\zeta_h^4}\,\bigg]\,\frac{1}{\zeta}\,\partial_{\zeta}F\hspace{6cm}\\ 
      +\frac{\mathfrak{w}}{\zeta_{h}^2 f^2}\left[2if(1-2c\zeta^2)\frac{\zeta^2}{\zeta_{h}^2}
      +\big\{4i(1-a_z)-\mathfrak{w}\big\}\frac{\zeta^6}{\zeta_{h}^6}+\frac{\mathfrak{w}}{a_z}\right]F=0,
      \end{split}
      \end{equation}
where $a_z\equiv\mathfrak{w^2}/(\mathfrak{w}^2-\mathfrak{q}^2 f)$. Now we solve this
equation using the multiscale perturbative method
      \begin{equation}\label{equation44}
      F_z=F_{z}^{\scriptscriptstyle{(0)}}+\mathfrak{w}F_z^{\scriptscriptstyle{(1)}}+
      \mathfrak{q}\,^2G_{z}^{\scriptscriptstyle{(1)}}+\mathfrak{q}\,^2\,
      \mathfrak{w}H_{z}^{\scriptscriptstyle{(1)}}+\cdots\,.
      \end{equation}
Imposing the incoming wave condition at the  horizon one finds, for the first two perturbative 
terms, 
      \begin{equation}\label{equation48}
      F_{z}^{\scriptscriptstyle{(0)}}=\mathscr{C}_{z},
      \end{equation}
      \begin{equation}\label{equation49}
      \begin{split}
       F_{z}^{\scriptscriptstyle{(1)}}=\frac{1}{2}i\,\mathscr{C}_{z}\bigg[\gamma
       +\frac{2 \mathfrak{q}^2}{c \zeta_h^2 \mathfrak{w}^2}
       \left\{1-e^{c\left(\zeta ^2-\zeta_h^2\right)}\right\}
       +e^{-2 c \zeta_h^2}\left\{\text{Ei}\left(c \zeta ^2+c \zeta_h^2\right)
       -\text{Ei}\left(2 c \zeta_h^2\right)\right\}\\
       -\text{Ei}\left(c{\zeta ^2}-c\zeta_h^2\right)+ 
       \ln\left(\frac{1}{2}c\zeta_{h}^{2}\,f\right)\bigg], 
   \end{split}
      \end{equation}
where $\gamma$ is the Euler constant, $\text{Ei}(u)\equiv-\int_{-u}^{\infty}dt\,t^{-1}e^{-t}$
is the exponential integral function and $\mathscr{C}_{z}$ is an arbitrary constant.
Then the longitudinal component of the electric field, $E_z$, takes the form 
      \begin{equation}\label{equation51}
      \begin{split}
      E^{\scriptscriptstyle{(-)}}_{z}= & \frac{1}{2}\mathscr{C}_{z}f^{-i\mathfrak{w}/4}
      \Bigg(2+\bigg[\gamma
       +\frac{2 \mathfrak{q}^2}{c \zeta_h^2 \mathfrak{w}^2}
       \left\{1-e^{c\left(\zeta ^2-\zeta_h^2\right)}\right\}-\text{Ei}\left(c{\zeta ^2}-c\zeta_h^2\right)\\
       & +e^{-2 c \zeta_h^2}\left\{\text{Ei}\left(c \zeta ^2+c \zeta_h^2\right)
       -\text{Ei}\left(2 c \zeta_h^2\right)\right\}+ 
       \ln\left(\frac{1}{2}c\zeta_{h}^{2}\,f\right)\bigg]i\mathfrak{w}+
       \mathcal O(\mathfrak{w}^2,\mathfrak{q}^2)\Bigg).
      \end{split}
      \end{equation}
The Dirichlet condition on the AdS boundary, $E^{\scriptscriptstyle{(-)}}_{z}(\zeta)\Big{|}_{\zeta=0}=0$,
implies that
    \begin{equation}\label{QNF}
    \omega=-i\frac{\big(1-e^{-c{\zeta}_h^2}\big)}{2 c\zeta_h} q^2+\mathcal O (\mathfrak{q}^3).
    \end{equation}
This is the fundamental quasinormal frequency for this problem.

It is interesting to see that this frequency coincides with the pole of the retarded Green's function
for the longitudinal component. In order to see this explicitly, we can use the series expansion
for the longitudinal field near the boundary 
    \begin{equation}\label{hidrodinamicoexpansao}
      E^{\scriptscriptstyle{(-)}}_{z}=1+\frac{i(1-e^{- c{\zeta}_h^2})}{2c\zeta_{h}^2\,\mathfrak{w}}
      \mathfrak{q}^2+\frac{ie^{-c{\zeta}_{h}^{2}}(\mathfrak{w}^2-\mathfrak{q}^2)}{2\mathfrak{w}}
      \left(\frac{\zeta}{\zeta_h}\right)^2+\cdots\,,
    \end{equation}
where above we have chosen $\mathscr{C}_z=1$.
Comparing this expression with eq. (\ref{equation31}) we find 
    \begin{align}
    \mathfrak{A}^{\scriptscriptstyle{(-)}}_z & =1+
    \frac{i(1-e^{-c{\zeta}_h^2})}{2c\zeta_{h}^2\,\mathfrak{w}}\mathfrak{q}^2+\cdots\,,\\
    \mathfrak{B}^{\scriptscriptstyle{(-)}}_z & =-\frac{i}{4\mathfrak{w}}e^{-c{\zeta}_h^2}
    \left[\mathfrak{q}^2-2\mathfrak{w}^2+e^{c{\zeta}_h^2}\left(\mathfrak{q}^2
    -2ic\zeta_h^2\mathfrak{w}\right)\right]+\cdots\,,
    \end{align}
where the dots denote terms of second or higher order in $\mathfrak{w}$ and $\mathfrak{q}$.
Thus, in the hydrodynamic limit, the longitudinal scalar function $\Pi^{L}(\omega,q)$
reduces to
    \begin{equation}
     \label{PiL}
     \Pi^{L}(\omega,q)=\frac{N_c^2 T}{16\pi}\,
     \frac{e^{-c/(\pi T)^2}(\omega^2- q^2)}{\left(i\omega-D q^2\right)}+\cdots
    \end{equation}
and the associated current-current correlation functions take the form
\begin{equation}\label{Green_hydrodynamic}
\frac{G^{\scriptscriptstyle{R}}_{tt}}{q^2}=-\frac{G^{\scriptscriptstyle{R}}_{tz}}{\omega q}=
-\frac{G^{\scriptscriptstyle{R}}_{zt}}{\omega q}=\frac{G^{\scriptscriptstyle{R}}_{zz}}{\omega^2}=
\frac{N_c^2 T}{16\pi}\frac{e^{-c/(\pi T)^2}}{\left(i\omega-D q^2\right)}+\cdots\,,
\end{equation}
with
\begin{equation}
D=\frac{\pi T}{2c}\left[1-e^{-c/(\pi T)^2}\right]. 
\end{equation}
This value of $D$ coincides with a result obtained in an early study
of black-hole electromagnetic perturbations in the soft-wall model \cite{Atmaja:2008mt}.
It is also in accordance with the result obtained in \cite{Policastro:2002se}
in the limit of vanishing $c$, and with the result of 
\cite{Kim:2010zg}, when the electric charge $Q$ of the black hole is zero.

The correlation functions (\ref{Green_hydrodynamic}) have a pole at
$\omega=-iDq^2$, that is the fundamental quasinormal frequency
obtained in eq. (\ref{QNF}). In the dual field theory, this pole
corresponds to charge diffusion with diffusion coefficient $D$.  
 
\subsubsection{Transverse perturbations }

The solution of equation (\ref{equation26}) for the transversal component
of the gauge field is determined in a similar way as in the longitudinal case.
Imposing the incoming wave condition at the horizon, we obtain the following
solution for $E^{\scriptscriptstyle{(-)}}_{\alpha}$ ($\alpha=x,y$) in the hydrodynamical limit: 
      \begin{equation}\label{equation321}
      \begin{split}
      E^{\scriptscriptstyle{(-)}}_{\alpha}=\frac{1}{4}\mathscr{C}_{\alpha}f^{-i\mathfrak{w}/4}
      \Bigg( & 4+\bigg[\gamma+e^{-2 c \zeta_h^2}\left\{\text{Ei}\left(c \zeta ^2+c \zeta_h^2\right)
       -\text{Ei}\left(2 c \zeta_h^2\right)\right\}\\
       & -\text{Ei}\left(c{\zeta ^2}-c\zeta_h^2\right)+
       \ln\left(\frac{1}{2}c\zeta_{h}^{2}\,f\right)\bigg]i\mathfrak{w}+
       \mathcal O(\mathfrak{w}^2,\mathfrak{q}^2)\Bigg).
      \end{split}
      \end{equation}
where the $\mathscr{C}_{\alpha}$'s are arbitrary constants.
The Dirichlet condition on the AdS boundary, $E^{\scriptscriptstyle{(-)}}_{\alpha}|_{\zeta=0}=0$,
leads to an algebraic equation that does not have solutions compatible with the hydrodynamic approximation
$\mathfrak{w},\,\mathfrak{q}\gg 1$. So, we conclude that $\Pi^T(\omega,q)$ does not present
poles in this regime.

We can check that the correlation functions $G^{\scriptscriptstyle{R}}_{xx}$
and $G^{\scriptscriptstyle{R}}_{yy}$ do not present poles in the hydrodynamic limit,
expanding eq. (\ref{equation321}) near the boundary 
    \begin{equation}\label{hidrodinamicoexpansaotrans}
     E^{\scriptscriptstyle{(-)}}_{\alpha}=1+\frac{1}{2}i\,e^{-c{\zeta}_h^2}
     \left(\frac{\zeta}{\zeta_h}\right)^{2}\mathfrak{w}+\cdots \,,
    \end{equation}
where we have chosen $\mathscr{C}_{\alpha}=1$. Comparing
this equation with the result (\ref{equation31}) one finds 
    \begin{equation}
     \mathfrak{A}^{\scriptscriptstyle{(-)}}_{\alpha}=1+\cdots\,, \qquad\quad 
      \mathfrak{B}^{\scriptscriptstyle{(-)}}_{\alpha}=-\frac{1}{2}c\zeta_h^2
      +\frac{1}{2}i\,e^{-c{\zeta}_h^2}\,\mathfrak{w}\cdots\,,
    \end{equation}
where the dots denote terms of second or higher order in $\mathfrak{w}$ and $\mathfrak{q}$.
Thus, the transversal scalar function $\Pi^{T}(\omega,q)$ reads
    \begin{equation}
     \Pi^{T}(\omega,q)=\frac{N_c^2 T}{16\pi}\left[\frac{c}{\pi T}+ie^{-c/(\pi T)^2}
     \omega\right]+\cdots\,,
    \end{equation}
which shows that the retarded correlation functions
$G^{\scriptscriptstyle{R}}_{xx}=G^{\scriptscriptstyle{R}}_{yy}=\Pi^{T}(\omega,q)$
indeed present no poles.

\subsection{Numerical solutions}

\subsubsection{Power series method}

In order to look for more general solutions valid not only in the hydrodynamical limit,
we now use numerical methods. Classical arguments tell us that black branes do not emit
radiation. Using this information, we only consider the incoming wave solutions at the horizon.
So, we write  $\psi_{j}=e^{-i\omega r_{*}}\varphi_{j}$ ($j=x,y,z$) in the Schr\"odinger like equations
\eqref{equation207} and \eqref{equation209} obtaining 
    \begin{equation}\label{equation216}
     P_j\,\frac{d^2\varphi_{j}}{d\zeta^2}+
    Q_j\,\frac{d\varphi_{j}}{d\zeta}+R_j\,\varphi_{j}=0,
    \end{equation}
where
\begin{align*}
P_j & =f\zeta^{2}({\omega}^{2}-{q}^{2}f)^{2},\qquad\qquad
Q_j=\zeta^{2}({\omega}^{2}-{q}^{2}f)^{2}\left\{\partial_{\zeta}f+2i\omega\right\},\\
R_j & =-({\omega}^{2}-{q}^{2}f)^{2}\frac{\zeta^2}{f}V_{_T}-4 q^2 \delta_{jz}
\left[q^2 f+c\zeta^2f(\omega^2-q^2 f)-\omega^2\left(1-3\frac{\zeta^4}{\zeta_{h}^4}\right)\right]
\left(\frac{\zeta}{\zeta_h}\right)^4,
\end{align*}
and $V_{_T}$ is the transverse potential (\ref{Vtransversal}).
We will solve equation (\ref{equation216}) numerically using the Horowitz-Hubeny method
\cite{Horowitz:1999jd}. Since the wave functions $\psi_{j}$ satisfy an incoming wave
condition at the horizon, the variables $\varphi_j$ tend to constants as $\zeta\rightarrow\zeta_h$.
Then we look for solutions of \eqref{equation216} in the form of a
simple power series near the horizon: 
      \begin{equation}\label{equation220}
      \varphi_{j}^{\scriptscriptstyle{(-)}}(\zeta)=\sum_{n=0}^{\infty}a_{nj}^{\scriptscriptstyle{(-)}}
      \left(1-\frac{\zeta}{\zeta_h}\right)^{n}\,,
      \end{equation}
where the first few coefficients of these series are shown in \eqref{equation409}. 
Substituting the foregoing expression in eq. (\ref{equation216})
one obtains a recurrence relation for the coefficients $a_{jn}^{\scriptscriptstyle{(-)}}$.
These coefficients depend on the frequency $(\widetilde{\omega})$, on the
wave number $(\widetilde{q})$ and the  temperature $(\widetilde{T})$ normalized
by the parameter $\sqrt{c}$. Note that in general the frequency is a complex number 
$\widetilde{\omega}=\widetilde{\omega}_{\scriptscriptstyle{R}}-i
\widetilde{\omega}_{\scriptscriptstyle{I}}$,
where the real part is related to the localization of the peak of the
corresponding spectral function and the imaginary part to the peak width.

The quasinormal mode solution must also satisfy the Dirichlet condition on the AdS
boundary, so that
       \begin{equation}\label{equation222}
      \sum_{n=0}^{\infty}a_{jn}^{\scriptscriptstyle{(-)}}
      (\widetilde{\omega},\,\widetilde{q},\,\widetilde{T})=0\quad\qquad (j=x,y,z).
      \end{equation}
The roots of these equations represent the frequencies of the quasinormal modes 
produced by the electromagnetic perturbations of the black hole \eqref{equation1}
in the AdS/QCD soft-wall model. Numerically, the infinite series \eqref{equation222}
are truncated at a finite order, and the order of truncation is chosen
adopting the criterion that the fractional difference between the roots of two
successive partial sums is smaller than a certain value.

\subsubsection{Breit-Wigner resonance method}

The power series method is suitable to find quasinormal frequencies
in the intermediate and high temperature regimes ($\widetilde{T}\gtrsim 1$).
However, its convergence gets poor at very low temperatures ($\widetilde{T}\ll 1$).
In this regime, the behavior of the effective potentials
$V_{\scriptscriptstyle{T}/\scriptscriptstyle{L}}(r_{\ast})$ enable us to use the Breit-Wigner approach. 
This method was originally developed to calculate scattering amplitudes in quantum mechanics.
Then it was used to study quasinormal modes of ultrarelativistic stars 
\cite{Thorne:1969,Chandrasekhar:1991II,ChandrasekharFerrari,Chandrasekhar:1992ey}.
Recently, it has been used to calculate quasinormal modes of AdS black holes 
\cite{Berti:2009wx,Miranda:2009uw}.

We take the normalizable solutions $\psi^{\scriptscriptstyle{(1)}}_j$ of the
Schr\"odinger like equations (\ref{equation207}) and (\ref{equation209})
that can be written as linear combinations of the near horizon solutions 
    \begin{equation}\label{equation223}
     \psi^{\scriptscriptstyle{(1)}}_j=\mathfrak{C}^{\scriptscriptstyle{(1)}}_j
     e^{-i\omega r_*}+\mathfrak{D}^{\scriptscriptstyle{(1)}}_j e^{+i\omega r_*}=
     \alpha_j(\omega)\cos (\omega r_*) - \beta_j(\omega) \sin (\omega r_*)\,.
    \end{equation}
At the quasinormal mode frequency $\mathfrak{C}^{\scriptscriptstyle{(1)}}_j=0$.
We analyse the behavior of the solutions in a neighborhood of these frequencies, where 
$\mathfrak{C}^{\scriptscriptstyle{(1)}}_j \sim (\omega-\omega_{\scriptscriptstyle{\text{QNM}}})$
with $\omega_{\scriptscriptstyle{\text{QNM}}}=\omega_{\scriptscriptstyle{R}}-
i\omega_{\scriptscriptstyle{I}}$. This implies that 
    \begin{equation}\label{equation224}
     \alpha_j^2+\beta_j^2=4\,\mathfrak{C}^{\scriptscriptstyle{(1)}}_j
     \,\mathfrak{D}^{\scriptscriptstyle{(1)}}_j\approx
     \text{const.}\times\left[(\omega-\omega_{\scriptscriptstyle{R}})^2+
     \omega_{\scriptscriptstyle{I}}^2\right].
    \end{equation}
The quasinormal frequencies are determined first minimizing this equation considering $\omega$ real.
Then the imaginary part can be obtained adjusting a parabola to eq. (\ref{equation224}). 
On the other hand the imaginary part can also be obtained imposing the incoming wave condition 
on the complex solutions of eqs. (\ref{equation207}) and (\ref{equation209}) with complex
frequency \cite{Chandrasekhar:1991II}:
    \begin{equation}\label{equationconplexa}
    \omega_{c}=\omega-i\,\omega_{\scriptscriptstyle{I}},\quad\qquad
    \psi_{jc}=\psi_j-i\,\psi_{j\scriptscriptstyle{I}}.
    \end{equation}
Substituting this complex solution in the differential equations (\ref{equation207})
and (\ref{equation209}) we find
    \begin{equation}\label{equationreal}
     \partial_{r_*}^2\psi_j-V\psi_j+(\omega^2-\omega_{\scriptscriptstyle{I}}^2)\psi_j
     -2\,\omega\,\omega_{\scriptscriptstyle{I}}\,\psi_{j\scriptscriptstyle{I}}=0,
    \end{equation}
    \begin{equation}\label{equationimaginario}
     \partial_{r_*}^2\psi_{j\scriptscriptstyle{I}}-V\psi_{j\scriptscriptstyle{I}}
     +(\omega^2-\omega_{\scriptscriptstyle{I}}^2)
     \psi_{j\scriptscriptstyle{I}}+2\,\omega\,\omega_{\scriptscriptstyle{I}}\,\psi_{j}=0,
    \end{equation}
where $V=V_{_T}$ for $j=x,y$ and $V=V_{_L}$ for $j=z$.
Considering that $\omega_I\ll\omega$, we find that the complex solution can be written in the form
    \begin{equation}
    \begin{split}
     \psi_{jc}=\frac{1}{2}[(\alpha_j+\omega_{\scriptscriptstyle{I}}
     \partial_{\omega}\beta_j)-i & (\omega_{\scriptscriptstyle{I}}
     \partial_{\omega}\alpha_j-\beta_j)]e^{i\omega r_*}\\
     & +\frac{1}{2}[(\alpha_j-\omega_{\scriptscriptstyle{I}}
     \partial_{\omega}\beta_j)-i(\omega_{\scriptscriptstyle{I}}
     \partial_{\omega}\alpha_j+\beta_j)]
     e^{-i\omega r_*}.
     \end{split}
     \end{equation}
Imposing the incoming wave condition at the horizon  one obtains the imaginary part of the frequency 
as
    \begin{equation}\label{equation225}
    \omega_{\scriptscriptstyle{I}}=-\frac{\alpha_j}{\partial_{\omega}\beta_j}=
    \frac{\beta_j}{\partial_{\omega}\alpha_j},
    \end{equation}
where $\partial_{\omega}$ is the derivative with respect to the frequency evaluated at 
$\omega=\omega_{\scriptscriptstyle{R}}$  \cite{Chandrasekhar:1991II} .

\subsubsection{Numerical results}

The quasinormal frequencies were determined as a function of the temperature,
for the case of zero wave number $q=0$. In the region of high temperatures,
the frequencies show a linear behavior that is in agreement with \cite{Kovtun:2005ev}.
In this region the infrared cutoff is negligible and the frequencies have the form
$\widetilde{\omega}^2_{\scriptscriptstyle{R/I}}=4(n+1)^2\widetilde{T}^2$ with $n=0,\,1,\,2,...\,$.

%
\FIGURE{
\centering\includegraphics[width=5.00cm,angle=-90]{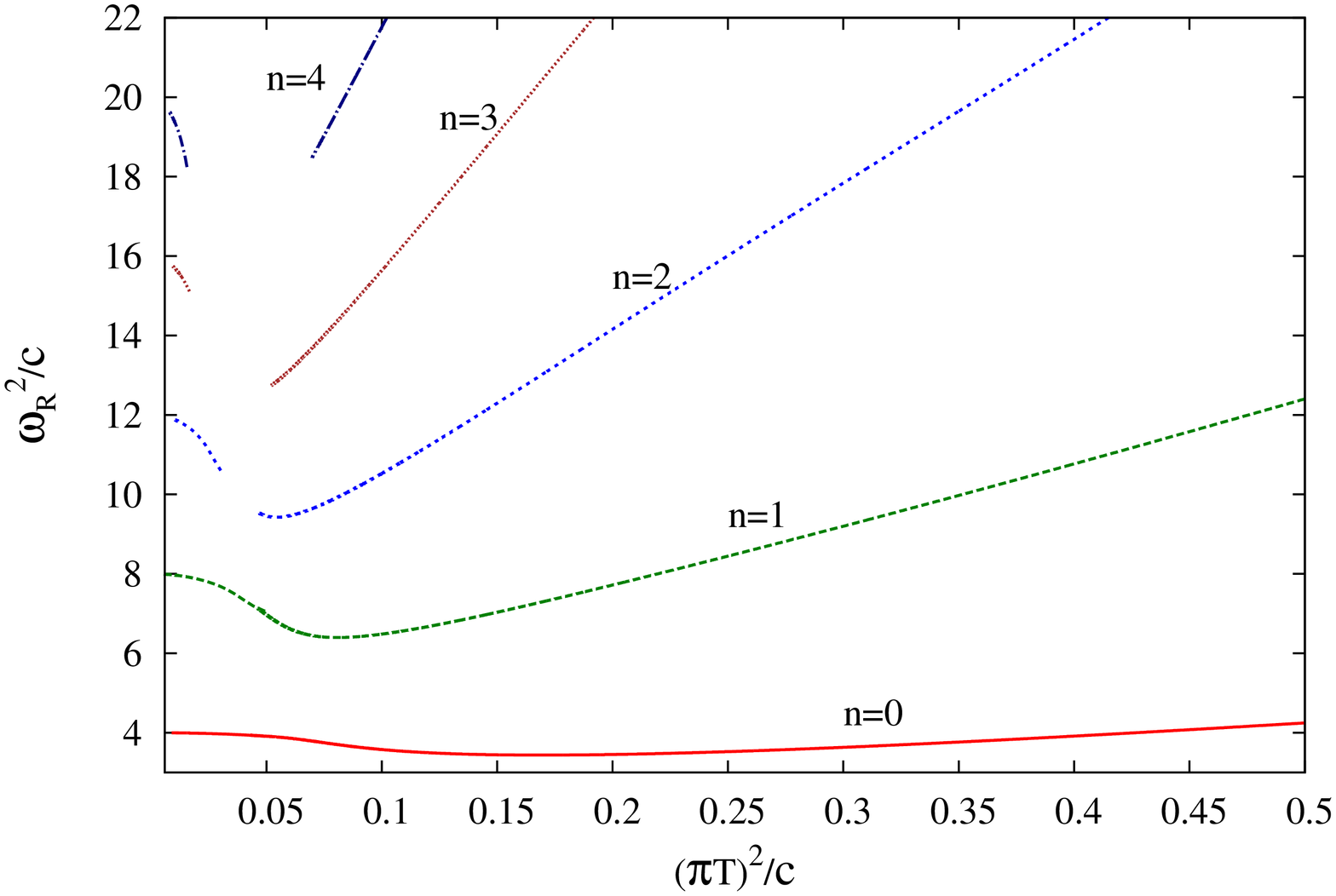}
\centering\includegraphics[width=5.00cm,angle=-90]{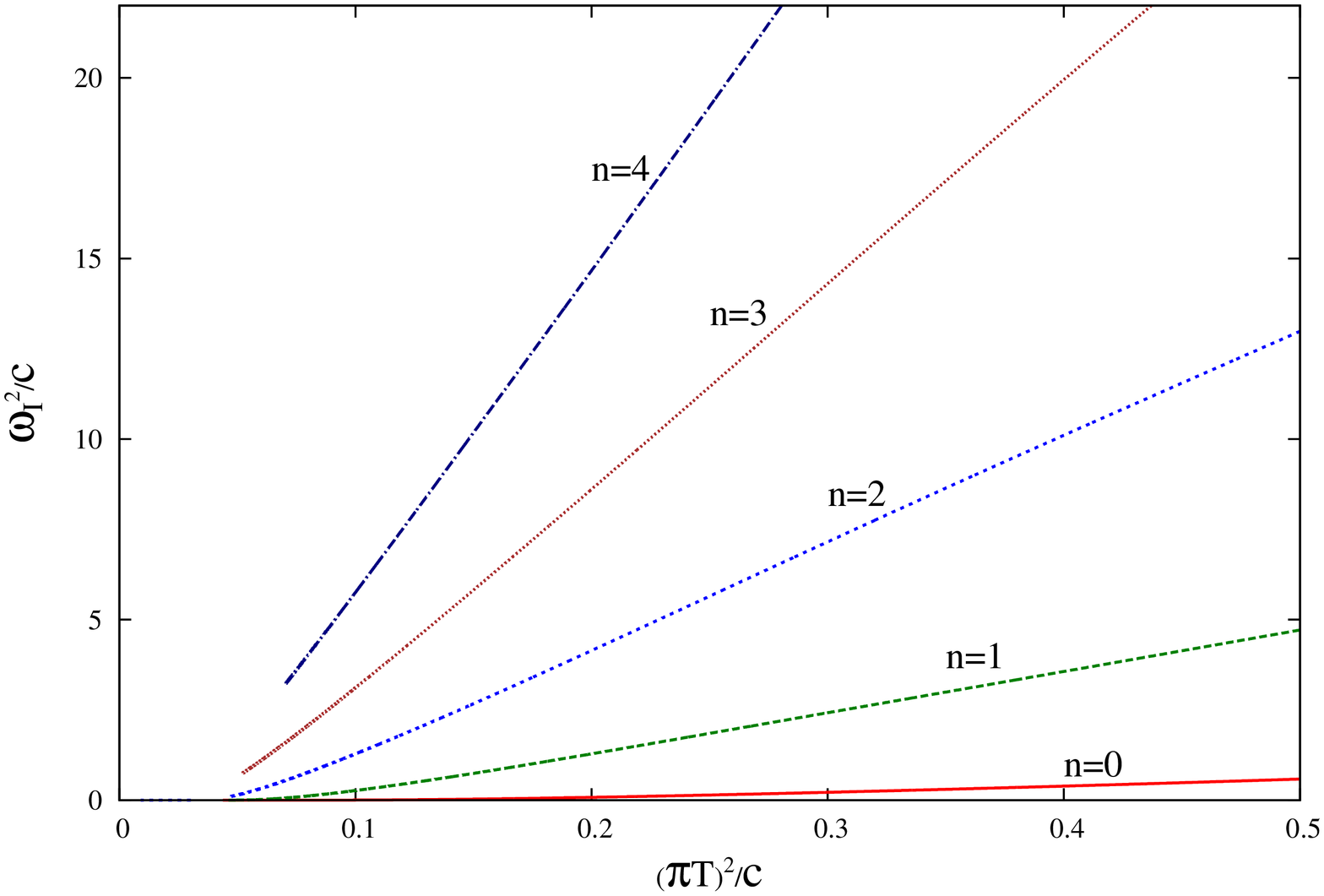}
\centering\caption{Numerical results for the quasinormal frequencies.
On the left panel we show the real part, associated with mass of the vector mesons.
On the right panel we show the imaginary part associated with the decay
time of the quasiparticle states.}
\label{modosquasenormais}
}

We show in figure \ref{modosquasenormais} the results for the square of the
real and of the imaginary parts of the frequencies in terms of $\widetilde{T}^2=(\pi T)^2/c$
for the first five modes $n=0,$ $1,\,...,\,4$ with $q=0$. 
The convergence of the power series method depends on the excitation number $n$.
For the first mode the power series method has a nice convergence up to the temperature 
$\widetilde{T}^2=0.0438$. For lower temperatures we used the Breit-Wigner resonance method. 
So, the two methods are complementary. This works well for the first two modes.
For the other excited states the convergence is poor for intermediate temperatures for both methods.
That is the reason why the curves are discontinuous. 

From figure \ref{modosquasenormais} one notes that in the zero temperature limit,
the real part of the QN frequencies coincide with the corresponding 
vector-meson mass spectrum of eq. (\ref{Espectrum}). One also notes that
at low temperature the behavior is not linear, thanks to the effect of the
infrared soft-wall cutoff.  Then, for the first two modes, increasing the
temperature $\widetilde{\omega}^2_{\scriptscriptstyle{R}}$ decrease until
they reach a minimum value at some critical temperatures. For higher temperatures,
$\widetilde{\omega}_{\scriptscriptstyle{R}}$ increases approaching a linear dependence
on the temperature. On the other hand, we observe in Fig.  \ref{modosquasenormais}
that the square of the imaginary part of the frequencies $\widetilde{\omega}^2_{\scriptscriptstyle{I}}$
increase monotonically with the square of the temperature approaching constant slopes. 
These results for the real and imaginary parts of the frequencies of the vector-meson
quasinormal modes are similar to what was found for scalar glueballs in Ref. \cite{Miranda:2009uw}.

\subsection{Dispersion relations}

In this subsection, we investigate the momentum dependence of the electromagnetic
quasinormal frequencies for both the longitudinal and transverse sectors.

\subsubsection{Longitudinal perturbations}

The longitudinal sector of perturbations is characterized by the presence of
a hydrodynamic quasinormal mode, which reduces to the diffusion mode \eqref{QNF} in the limit of small
frequencies and wavenumbers in comparison to the temperature. Figure \ref{HidrodinamicoMaio}
show the analytical approximation (\ref{QNF}) and the numerical results for
the dispersion relation of this hydrodynamic QNM.

\FIGURE{
\centering\includegraphics[width=7cm,angle=-90]{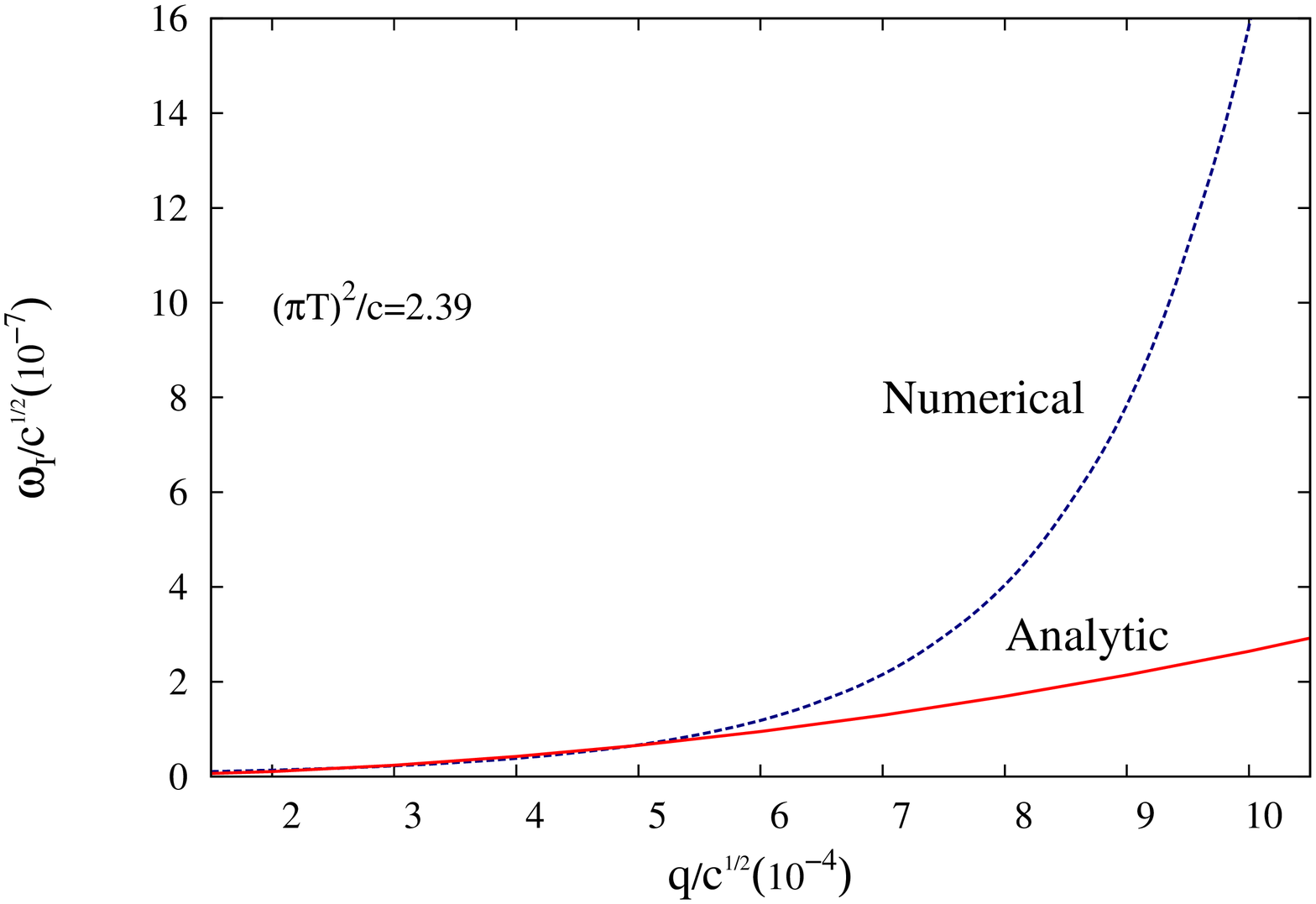}
\centering\caption{The dispersion relation of the hydrodynamic QNM for a fixed value of
    temperature $(\pi T)^2/c=2.39$. As shown in the last subsection,
    this kind of quasinormal mode appears only in the longitudinal sector of perturbations.}
    \label{HidrodinamicoMaio}
}

The variation of the non-hydrodynamic QNM frequencies with the wavenumber is shown in Fig.
\ref{longitudinalrelations} for $\widetilde{q}\leq 0.8$. We do not present the results
for higher values of wavenumber, because the convergence of the methods is very poor
for $\widetilde{q}>0.8$. As one can see, the real part
of the frequencies increases with $\widetilde{q}$. Such a behavior is consistent with
the results shown in figure \ref{espectral1}, where the position of the peaks of the
spectral function shifts to higher energies when the wavenumber value increases.
A similar thing happens with the imaginary part of the quasinormal frequencies
(width of the peaks), which increases with $\widetilde{q}$ for small values of temperature
($\widetilde{T}^2=0.0708$). In the intermediate- and high-temperature regimes,
the spectral functions do not present peaks, and so the quasinormal frequencies cannot
be associated with quasiparticle excitations in the plasma.
As the plasma gets hotter ($\widetilde{T}^2=0.2$), the behavior of the imaginary part
of the frequencies change, and $\widetilde{\omega}^2_{\scriptscriptstyle{I}}$ decreases 
with $\widetilde{q}$. These results are in accordance with previous works
\cite{Kovtun:2005ev,Nunez:2003eq} on the electromagnetic perturbations of
AdS black holes without the presence of the dilaton $\Phi(\zeta)=c\zeta^2$.
\FIGURE{
\centering\includegraphics[width=5.00cm,angle=-90]{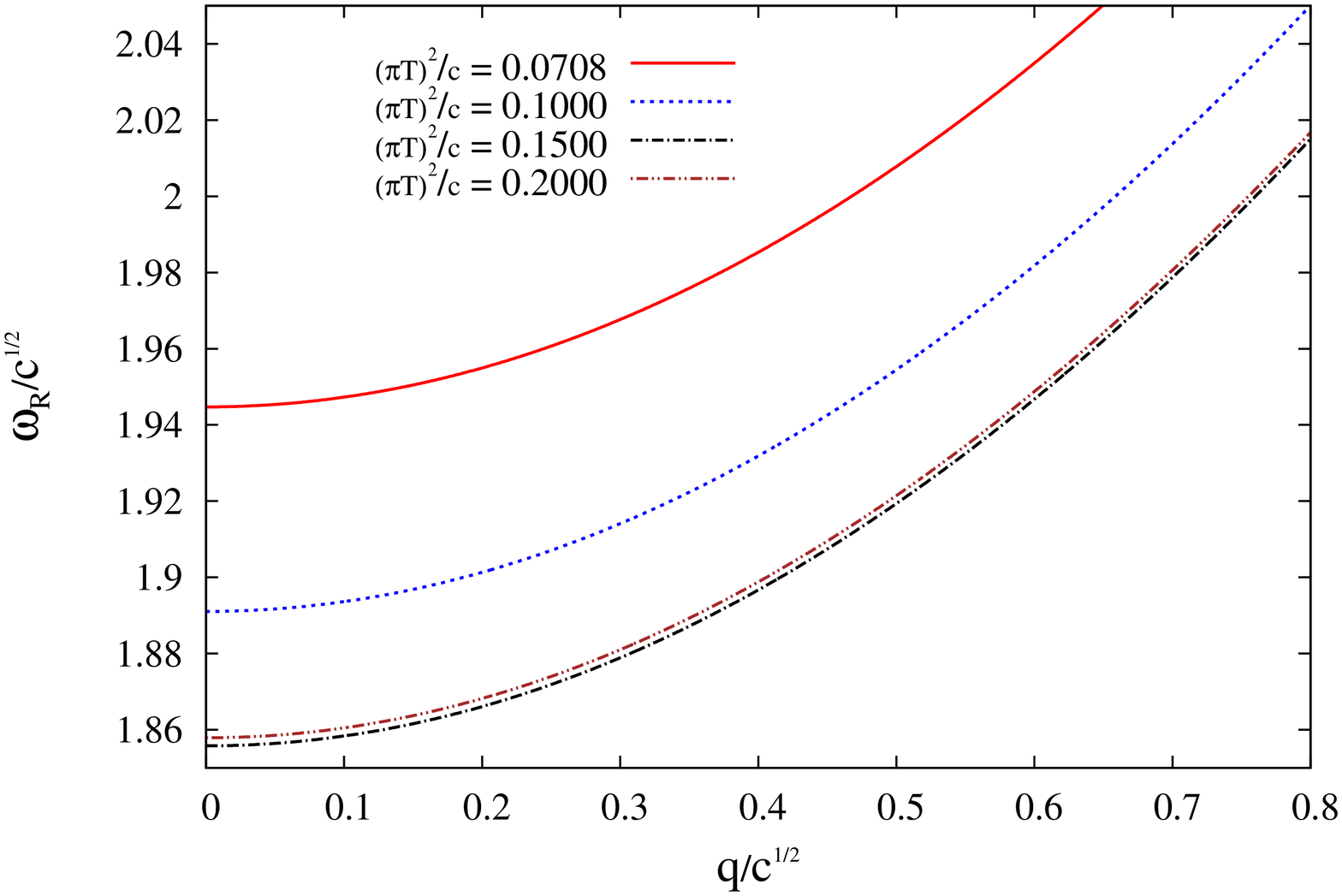}
\centering\includegraphics[width=5.00cm,angle=-90]{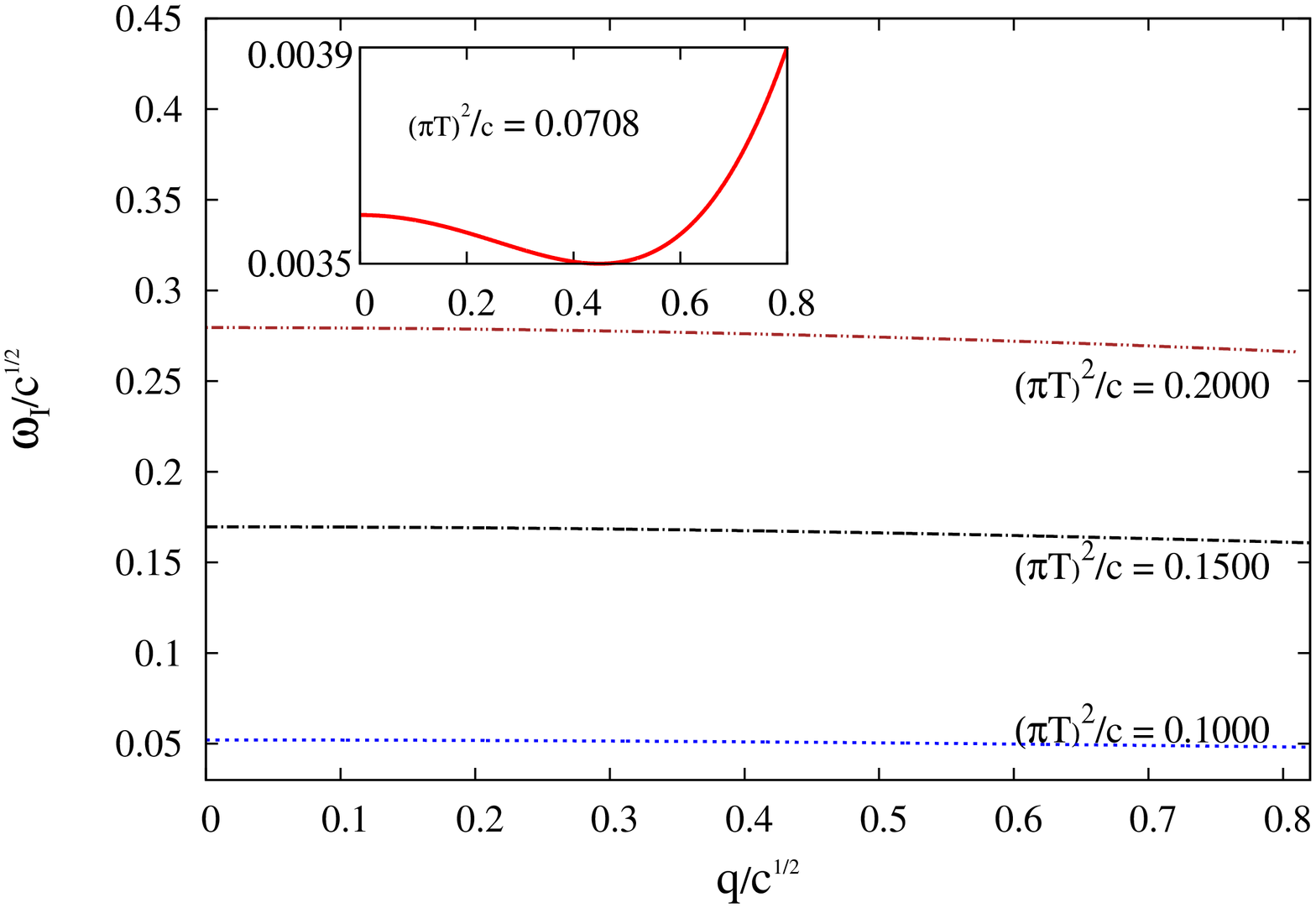}
\centering\caption{Numerical results for the dispersion relations of the first non-hydrodynamic
    quasinormal mode of longitudinal perturbations. The real part of the frequencies
    is shown on the left side, and the imaginary part on the right side.}
    \label{longitudinalrelations}
}

Some selected numerical results for the quasinormal frequencies are shown in Table \ref{dispersaolong}.
We also list in this table some values of the quantity
\begin{equation}\label{variacao}
\Delta_{q}\equiv\frac{\big|{\omega}_{\scriptscriptstyle{R}}(q)-
\sqrt{({\omega}_{\scriptscriptstyle{R}}^{0})^2+q^2}
\big|}{\sqrt{({\omega}_{\scriptscriptstyle{R}}^{0})^2+q^2}},
\end{equation}
which is a measure of the fractional difference between the real part
of the QNM frequencies and the results obtained from the relativistic
dispersion relation $\omega=\sqrt{({\omega}_{\scriptscriptstyle{R}}^{0})^2+q^2}$,
where ${\omega}_{\scriptscriptstyle{R}}^{0}\equiv
{\omega}_{\scriptscriptstyle{R}}(q)|_{q=0}$ can be regarded
as the vector-meson mass for very low temperatures.
\TABLE{
\centering
\begin{tabular}{cccc|ccc}
      &	 \multicolumn{3}{c}{$\widetilde{q}=0.05$} & \multicolumn{3}{c}{$\widetilde{q}=0.8$} \\ \cline{2-7}
  $\widetilde{T}^2$    & $\widetilde{\omega}_{\scriptscriptstyle{R}}$ &
  $\widetilde{\omega}_{\scriptscriptstyle{I}}$ & $\Delta_{q}$ & $\widetilde{\omega}_{\scriptscriptstyle{R}}$
  & $\widetilde{\omega}_{\scriptscriptstyle{I}}$ & $\Delta_{q}$\\ \hline
  0.0708 & 1.94536 & 0.00365 & $1.20416\times10^{-6}$ & 2.10249 & 0.00388 & $1.67812\times10^{-4}$ \\
  0.1000 & 1.89170 & 0.05207 & $9.58394\times10^{-6}$ & 2.05006 & 0.04825 & $1.58542\times10^{-3}$ \\
  0.1500 & 1.85643 & 0.16959 & $1.50934\times10^{-5}$ & 2.01499 & 0.16118 & $2.90839\times10^{-3}$ \\
  0.2000 & 1.85855 & 0.27943 & $1.46131\times10^{-5}$ & 2.01677 & 0.26643 & $2.99261\times10^{-3}$ \\ \hline
 \end{tabular} 
\centering\caption{Numerical results of the first non-hydrodynamic longitudinal QNM for
$\widetilde{q}=0.05$ and $\widetilde{q}=0.8$ and selected values of temperature.}
\label{dispersaolong}
}

\subsubsection{Transverse perturbations}

We show in figure \ref{transvesalrelations} the numerical results obtained
for the quasinormal frequencies of transverse perturbations in function
of the wavenumber $\widetilde{q}$. On one hand, the real part of the QNM frequencies
have a similar behavior to that of the longitudinal sector. On the other hand,
(the negative of) the imaginary part of $\widetilde{\omega}$ increases with the
momentum for small temperatures and small values of $\widetilde{q}$. 
In the intermediate- and high-temperature regime, $\widetilde{\omega}_{\scriptscriptstyle{I}}$
decreases with the wavenumber (see Fig. \ref{transvesalrelationshigh}). These results resemble
those for the QNM spectrum associated to scalar glueballs in the soft-wall model \cite{Miranda:2009uw}.

Some numerical results for $\widetilde{\omega}=\widetilde{\omega}_{\scriptscriptstyle{R}}-i\,
\widetilde{\omega}_{\scriptscriptstyle{I}}$ and for the fractional difference of
$\widetilde{\omega}_{\scriptscriptstyle{R}}$ with respect to the 
relation (\ref{variacao}) are listed in Table \ref{dispersaotrans}. As expected,
the value of $\Delta_{q}$ increases with the momentum and goes to zero in the limit of zero temperature.
In fact, the non-zero value of $\Delta_{q}$, indicating a departure of the standard relativistic
dispersion relation, is a finite-temperature effect.
%
\FIGURE{
\centering\includegraphics[width=5.00cm,angle=-90]{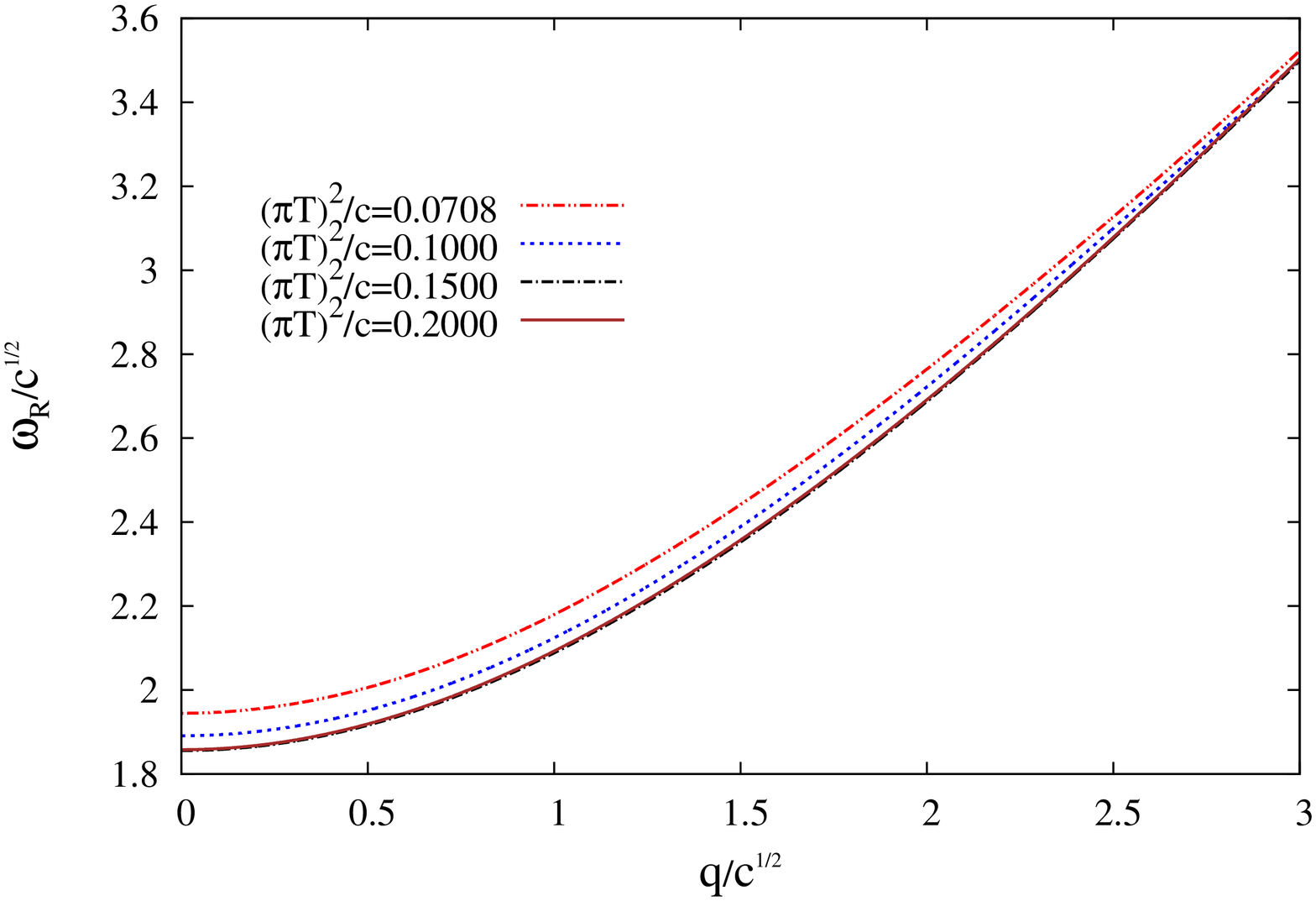}
\centering\includegraphics[width=5.00cm,angle=-90]{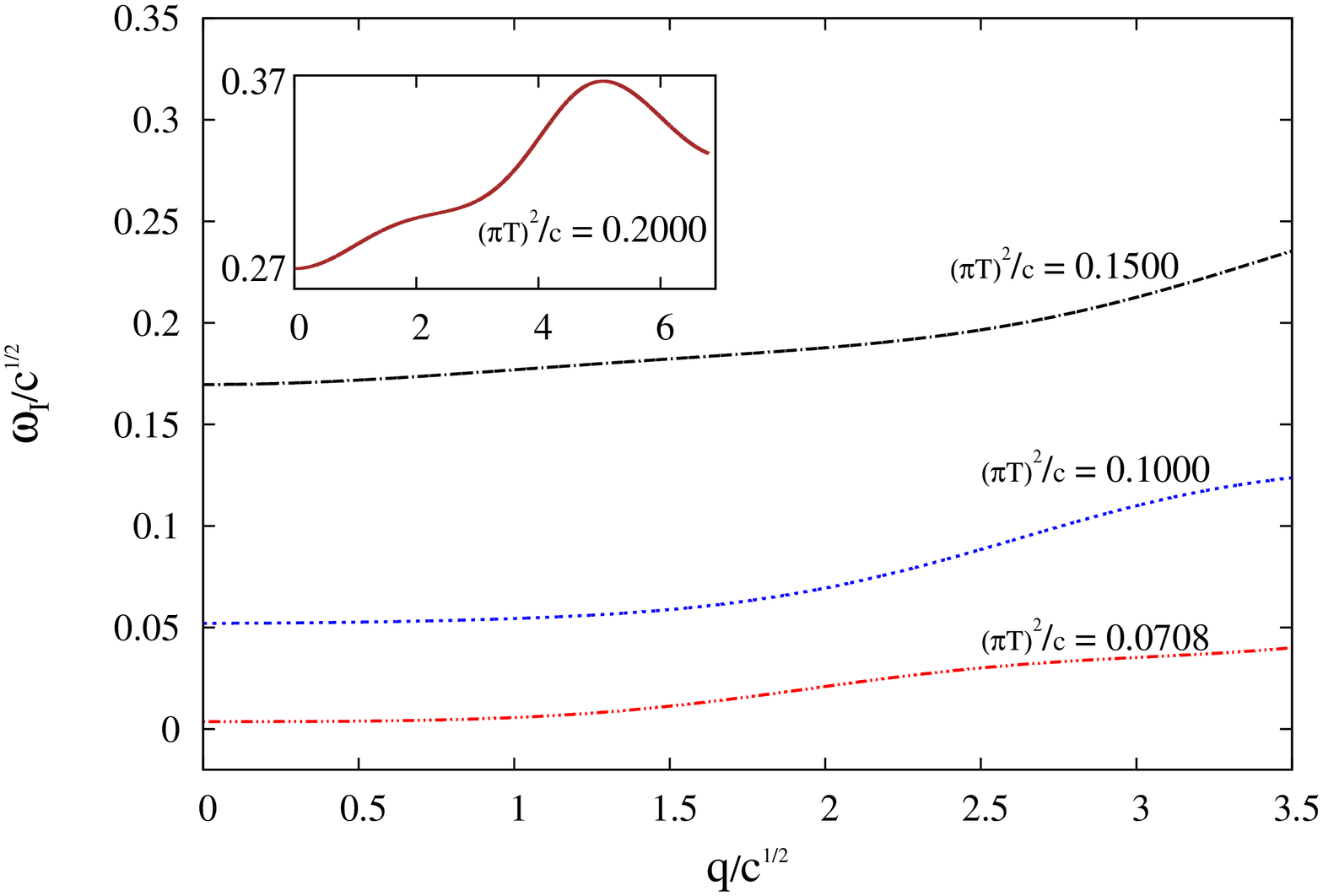}
\centering\caption{The QNM dispersion relations for transverse perturbations and selected values
    of $\widetilde{T}^2$ in the low temperature regime.}
    \label{transvesalrelations}
}

%
\FIGURE{
\centering\includegraphics[width=5.00cm,angle=-90]{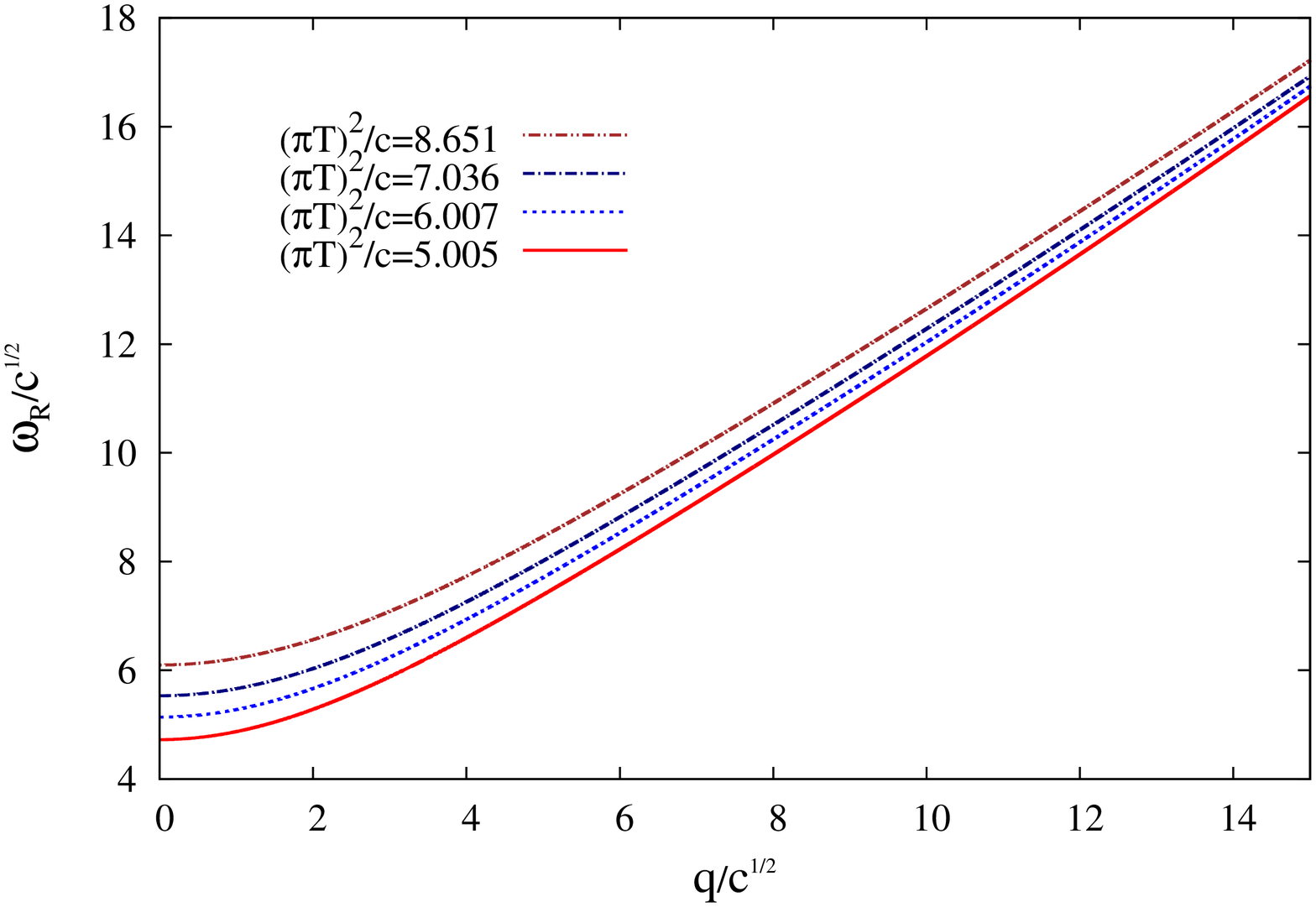}
\centering \includegraphics[width=5.00cm,angle=-90]{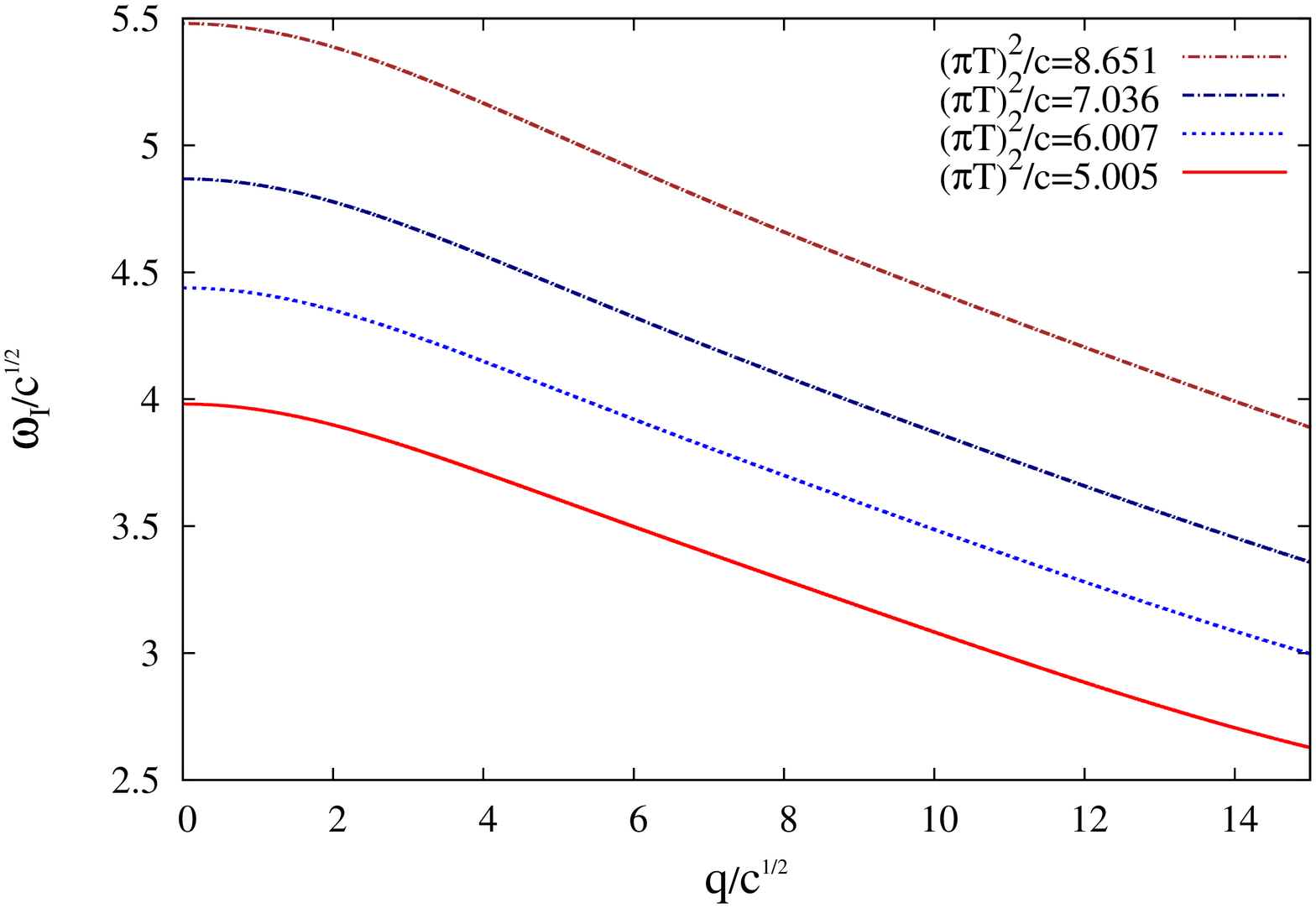}
\centering\caption{The QNM dispersion relations for transverse perturbations and selected values
    of $\widetilde{T}^2$ in the high temperature regime.}
    \label{transvesalrelationshigh}
}
\TABLE{
\centering
\begin{tabular}{cccc|ccc}
      &	 \multicolumn{3}{c}{$\widetilde{q}=0.05$} & \multicolumn{3}{c}{$\widetilde{q}=2.5$} \\ \cline{2-7}
  $\widetilde{T}^2$    & $\widetilde{\omega}_{\scriptscriptstyle{R}}$ & $\widetilde{\omega}_{\scriptscriptstyle{I}}$
  & $\Delta_q$ & $\widetilde{\omega}_{\scriptscriptstyle{R}}$ & $\widetilde{\omega}_{\scriptscriptstyle{I}}$
  & $\Delta_q$\\ \hline
  0.0708	& 1.94534 & 0.00365 & $1.09887\times10^{-5}$ & 3.12704 & 0.03011 & $1.27201\times10^{-2}$ \\
  0.2000	& 1.85853 & 0.27952 & $2.78399\times10^{-5}$ & 3.07811 & 0.30639 & $1.17724\times10^{-2}$ \\
  5.0050	& 4.72565 & 3.98104 & $2.37394\times10^{-5}$ & 5.56312 & 3.85751 & $4.06406\times10^{-2}$ \\
  8.6510	& 6.09834 & 5.48010 & $1.68570\times10^{-5}$ & 6.80566 & 5.34037 & $3.26327\times10^{-2}$ \\  
  \hline
\end{tabular} 
\centering\caption{Numerical results of the fundamental transverse QNM for $\widetilde{q}=0.05$
and $\widetilde{q}=2.5$ and selected values of temperature.}
\label{dispersaotrans}
}

\section{Conclusions} 

In this article we studied the thermal behavior of vector mesons in the soft-wall AdS/QCD model.
This was done, considering gauge fields in an AdS black hole space with a scalar background field
representing a smooth infrared cut-off. The equations of motion for longitudinal and transverse
components were obtained and expressed in Schr\"odinger like form. The analysis of the
corresponding potentials indicates the existence of quasiparticle states of the vector mesons.
At high temperatures and zero wavenumber, the potentials have the shape of an infinite barrier.
In this regime there are no quasiparticle  states in the dual theory. For smaller values of the
temperature, below $\widetilde{T}^2_c =  0.538$ the potential presents a well. In this case,
there are quasiparticle states in the dual theory, corresponding to vector mesons at finite temperature. 

We computed the retarded Green's functions. The imaginary part of these correlation functions,
which are known as the spectral functions, contains information about the formation of quasiparticle
states. As expected, it was found that quasiparticle states are formed at low temperatures where the potential 
has a well. As the temperature decreases, the quasiparticle states are more stable since the width
of the peaks of the spectral function shrink. The thermal masses of the states are given by the frequency
values of the peaks. We also analysed the case of non vanishing wave number and found that the quasiparticle
states are more unstable in this case, since the width of the peaks increase with the wave number. 

We also studied the frequencies of the quasinormal modes of the gauge fields in the finite temperature
soft-wall model. In the hydrodynamical regime we obtained analytically the quasinormal frequencies
for the longitudinal component. The corresponding charge transport coefficient was also found. Outside
the hydrodynamical limit, we employed two numerical methods: power series and Breit-Wigner resonance method. 
As expected, the quasinormal frequencies correspond to the poles of the current-current correlation functions.

The QNM dispersion relations show that the real part of the frequencies always increases with the momentum.
For small values of temperature and wavenumber, the behavior of the real part of the frequencies
is similar to that of the relativistic dispersion relation $\omega=\sqrt{({\omega}_{\scriptscriptstyle{R}}^{0})^2+q^2}$,
for both sectors of perturbations (longitudinal and transverse), as one can see in Tables \ref{dispersaolong} and
\ref{dispersaotrans}. The imaginary part of the first non-hydrodynamic QNM frequency increases
with the momentum for very small values of $\widetilde{T}$ and $\widetilde{q}$, while in the intermediate
and high temperature regimes $\widetilde{\omega}_{\scriptscriptstyle{I}}$ is a monotonically decreasing
function of $\widetilde{q}$.


\section*{Acknowledgements}

We thank the Centro de Armazenamento de Dados e Computa\c{c}\~ao Avan\c{c}ada of the
Universidade Estadual de Santa Cruz for the computational support. The authors are partially
supported by the Conselho Nacional de Desenvolvimento Cient\'ifico e Tecnol\'ogico (CNPq)
and the Coordena\c{c}\~ao de Aperfei\c{c}oamento de Pessoal de N\'ivel Superior (Capes),
Brazilian agencies.



\end{document}